%% file: conselice.tex
\documentstyle[10pt,aaspp4] {article}

\def\deg{$^{\circ}\,$}
\def\solm{M$_{\odot}\,$}
\def\kms{km~s$^{-1}\,$}

\begin{document}

\title{Galaxy Populations and Evolution in Clusters I:  Dynamics and
the Origin of Low-Mass Galaxies in the Virgo Cluster}

\author{Christopher J. Conselice$^{1,2}$, John S. Gallagher III$^{1}$}

\affil{Department of Astronomy, University of Wisconsin-Madison}

\author{Rosemary F.G. Wyse}

\affil{Department of Physics \& Astronomy, The Johns Hopkins University}

\altaffiltext{1}{Visiting Astronomer, Kitt Peak National Observatory,
National Optical Astronomy Observatories, which is operated by the
Association of Universities for Research in Astronomy, Inc. (AURA) under
cooperative agreement with the National Science Foundation. }

\altaffiltext{2}{Also: Space Telescope Science Institute, 3700 San Martin
Drive. Baltimore MD. 21218}
\begin{center}

{\it Accepted to the Astrophysical Journal}

\end{center}

\begin{abstract}

Early-type dwarfs are the most common galaxy in the local universe, yet
their origin and evolution remain a mystery.  Various cosmological 
scenarios predict that dwarf-like galaxies in 
dense areas are the first to form and hence should be the oldest stellar 
systems in clusters. By using radial velocities of early-type dwarfs
in the Virgo cluster we demonstrate that these galaxies are not an old
cluster population but have signatures of production from the infall
of field galaxies.  Evidence of this
includes the combined large dispersions and sub-structure in spatial and 
kinematic distributions for Virgo early-type dwarfs, and a velocity 
dispersion ratio with giant ellipticals expected for virialized and accreted
populations.  We also argue that these galaxies cannot originate from
accreted field dwarfs, but must have physically 
evolved from a precursor population, of different morphology, that fell into 
Virgo some time in the past.

\end{abstract}

\keywords{Galaxy Clusters: Virgo, Galaxies: Dwarf Ellipticals/Spheroidals, 
Galaxies: Formation, Evolution}

\section{Introduction}

    Early-type dwarfs are the most common galaxies in the
local universe, yet we know very little about them (cf. Gallagher \& Wyse 1994;
Ferguson \& Binggeli 1994).  This is due to
their very low luminosities and surface brightnesses hindering
any detailed studies.  Luminosity
functions of nearby clusters show that faint galaxies\footnote{For the 
purposes of this paper, we 
refer to these objects as dwarfs.  We further subdivide dwarfs into 
dwarf ellipticals (dE) and irregulars (dIrr).  Dwarf spheroidals are not 
used as a separate term in this paper but are included in the dwarf elliptical 
class.  The differences between dE's and dIrr will be discussed in \S 
2.1} are by far
more common than brighter galaxies (e.g., de Propris et al. 1995; 
Secker, Harris \& Plummer 1997; Trentham 1997; Phillipps et al. 1998).  
Additionally, in comparison to the field, galaxy clusters have steeper 
luminosity functions dominated by dE-like galaxies
(e.g., Ferguson \& Sandage 1989; de Propris et al. 1995
Secker et al. 1997;  Marzke \& da Costa 1997; Phillipps et al. 1998; 
Mobasher \& Trentham 1998).  How these objects originated, 
and continue to exist in large numbers in clusters is presently a mystery.

   A very basic question to ask is: where did these cluster dwarf galaxies
come from? Do they have an origin similar to the dwarf elliptical and 
spheroidal galaxies seen in groups such as the Local Group (cf. Mateo 1998; 
van den Bergh 2000)?  One basic difference between group and cluster dwarfs is 
the way they are
distributed.  In groups of galaxies, dwarf spheroidals are for the most
part clustered around giants (Mateo 1998; Grebel 1999). However, in 
clusters dwarfs
are in general $\it not$ distributed around the giants, but follow the 
general potential (e.g., Binggeli, Tarenghi \& Sandage 1990; Ferguson 1992;
Secker et al. 1997). In other respects, such as surface brightness, magnitude,
and color, bright dEs in clusters have properties similar to bright
Local Group dEs (Bothun et al. 1985; James 1991; Held \& Mould 1994; 
Phillipps et al. 1998).

  Dwarfs in rich clusters are faint and hard to identify and as a result they
were not discovered until Reeves (1956, 1977) and Hodge (1959, 1960, 1965) 
studied them in the Virgo and Fornax clusters.    Since then 
many observations, conducted at various wavelengths and employing 
various techniques, have been applied to cluster
dwarfs, with particular activity in the 1980s.   Many
of these studies were mainly interested in determining the galaxy luminosity
functions in clusters (Ferguson \& Sandage 1988; Phillipps et al. 1998; 
Mobasher \& Trentham 1998; Trentham 1998) and not necessarily in the 
properties of the dwarf galaxies themselves.  Luminosity functions are 
important for cosmology and galaxy formation theories, but they
also reveal what the 
relative number densities of cluster dwarfs are in comparison with the field.  

There are at least four different ideas for explaining where
dEs seen in Virgo and other galaxy clusters originate.  These
are: (1) dEs are old primordial galaxies that formed early in the
Virgo proto-cluster (e.g., White \& Frenk 1991; 
Ostriker 1993; Farooqui 1994) and have survived in the cluster environment
for several Gyrs. (2) dEs are the passively evolved 
compact blue and emission line galaxies seen at redshifts $z \sim 1$
(e.g., Babul \& Rees 1992; Koo et al. 1995; Koo et al. 1997) (3) dEs are
stripped and `harassed' spirals accreted into clusters,
creating a morphological transformation: spiral $\rightarrow$ dE (Mao \&
Mo 1998; Moore et al. 1998; Moore et al. 1999) (4) dEs evolved from dwarf 
irregulars (Lin \& Faber 1983).  Alternatively dwarf ellipticals may form 
though a combination of these scenarios.

Solution (4) was considered quite seriously and given
much attention in the 1980s, but has since been largely discarded (e.g.,
Bothun et al. 1986; Gallagher \& Hunter 1989) due to the problem of 
explaining large populations of dEs that are more massive, brighter, and 
more compact than typical dwarf irregulars. Since any evolution from 
dIrr to 
dE requires fading, bright dEs cannot be accounted for by this evolution.

Possibility (2) was inspired by the large numbers of faint blue galaxies
seen in redshift surveys (Babul \& Rees 1992).
These blue galaxies have low masses, similar to masses of dwarfs 
(Koo et al. 1995). If no further star formation occurs then the 
magnitudes and surface brightnesses
of these objects would be similar to those of dwarf ellipticals 
(Koo et al. 1995), thus allowing for the possibility of
an evolutionary connection.
However, there are several potential problems with this idea, the most 
important being the fact that many Local Group dwarf ellipticals contain 
several episodes of star formation (cf. Mateo 1998) and hence have not
evolved passively.   A second problem is that the faint blue
galaxies are a field population, yet in the nearby universe most
dEs are in clusters.  Third, some
Virgo dEs are brighter than the luminosity to which these high redshift,
star-forming galaxies 
would evolve with no further star formation. Thus, a strictly 
passive evolution from a burst of star formation at $z \sim 1$ to the
present cannot account for cluster dEs.  

In this paper, we mainly address, 
and try to distinguish between, possibilities (1) and (3).
The hypothesis we test is that cluster dEs form either by the galaxy 
harassment scenario (e.g., Moore et al. 1998), or are an
old cluster population.  These models depend upon a very 
active source of accreted galaxies to account for the large population of
dEs in clusters. For the harassment idea to
be correct, there must be signatures that dEs were accreted and 
underwent mass stripping, or that a large population of spirals existed in 
the proto-cluster.  Modern Virgo spirals
have properties that suggest they have recently been accreted (Huchra
1984) and
this process likely occurred in the past as well.
Harassment has also been used to explain the morphological transformation
of
Butcher-Oemler galaxies seen at moderate cluster redshifts (z$\approx$0.8) into
modern dwarf galaxies (Moore et al. 1996).    A galaxy in a cluster, such as 
Virgo, interacts weakly 
with the other cluster members due to the high relative
velocities between systems.  Stars in the outer parts of accreted galaxies
are stripped
away due to the increase of internal energy from impulsive
interactions and at the same time gas is stripped by ram pressure 
(Lee, Richer, \& McCall 2000; Mori \& Burkert 2000).   
Over the 1 Gyr orbital time the 
galaxy begins to lose its stars and gas, and over a longer time can 
become a spheroidal system (Moore et al. 1998).  Before a galaxy is stripped
any gas it contains can become compressed by cluster tides producing a 
starburst.  Harassment does not differentiate between accreted galaxies and 
cluster members or galaxies on isotropic or anisotropic orbits.
This theory predicts that the larger spiral galaxies become S0s (Moore
et al. 1999) while the smaller ones will eventually resemble dwarf 
ellipticals.   It remains to be determined,
but seems unlikely, whether it is possible that each dE originated from a 
spiral.  The increase in number of star forming galaxies in z$\sim 1$ 
clusters may not need an 
explanation from an increase in the infall rate, since the field is 
rich in starbursts at this same redshift (e.g., Sullivan et al. 2000).

One way to test if cluster dEs originated from a relatively recently
accreted population
is to determine how their velocity and spatial
distributions compare to those of giant ellipticals
and spirals.  The one dimensional velocity dispersions
of clusters are typically less than 1000 \kms (for Virgo the value
is about 700 \kms).  From radial velocity studies of Virgo galaxies it
is known that the dispersion of the spiral
population is higher and has a flatter distribution (less relaxed)
than that of the ellipticals which is Gaussian, an expected form for 
a relaxed population (e.g., Huchra 1985, Schindler, Binggeli, \& 
B$\ddot{{\rm o}}$hringer 1999 (SBB99)).  This
velocity distribution difference is usually interpreted as a consequence of 
the infall of spiral galaxies.  As a very basic test we can
measure the velocity and spatial distributions of Virgo dwarf galaxies to 
determine if they have such accretion signatures, or instead signatures of an 
older relaxed cluster population.  Furthermore, we 
can compare these  
distributions to various predictions, including the effects of 
relaxation, virialization, equipartition and dynamical friction.   

 We show
that the early type dwarfs in Virgo, as well as all Virgo
galaxies except ellipticals,  have no strong signs of relaxation. 
Based on both their velocity and spatial distributions, we demonstrate that 
most Virgo dEs cannot have existed in Virgo for as long as the giant 
ellipticals have.
The kinematic properties of Virgo dwarf ellipticals also suggest that these 
galaxies originate from an accreted component, whose infall rate peaked 
some time in the past.
This paper is organized as follows: \S 2 describes the observations and
data, \S 3 gives our results, \S 4 describes various dynamical 
models,  \S 5 is a discussion of our results, and finally \S 6 
gives a summary.  A Virgo cluster distance of 18 Mpc is used throughout,
where 6\deg = 1.9 Mpc.  

\section{Data and Observations}

The Virgo Cluster of galaxies is the nearest rich cluster, 
presenting both advantages and problems for studying its 
galaxies and structure. It is close enough so that its galaxies can be 
studied in detail, but because it is so close, it is spread out over 10
degrees in the sky, making multiplexed observations of its galaxies 
time consuming.

Measuring radial velocities of faint, low surface brightness
dwarf galaxies is also technically very challenging.  The brightest dwarfs
can be studied with little difficulty, but the fainter ones, with
surface brightnesses significantly fainter than the night sky, though
very numerous in Virgo, are difficult to observe spectroscopically.  As such
we can  only determine
radial velocities of relatively bright dEs.
Previous to this work, the last published radial velocity catalog of
Virgo contained 91 dE + dS0 galaxies (Binggeli, Popescu \& Tammann 1993).
We increase the total number to 142 by measuring new velocities and combining
these with previously published values. 

We undertook radial velocity observations of selected Virgo 
galaxies with the Hydra multi-object spectrograph on the WIYN
3.5m telescope, and with the RC long-slit spectrograph  
on the Kitt Peak National Observatory 4m telescope.  

Hydra contained 95 fibers at the time these observations were made in 1999 
April and 2000 March.  For these observations Hydra's fibers fed the bench 
spectrograph with a 400 lines/mm grating during 1999 April and the 
600 lines/mm grating in 2000 March. Total dispersions of 2.1 \AA\, 
pixel$^{-1}$ and 1.4 \AA\,pixel$^{-1}$
and equivalent resolutions of 6.9\,\AA\, and 4.6\,\AA\, were obtained for the 
two gratings respectively, giving a typical error of 35 - 50 km s$^{-1}$ in
radial velocity. 
The wavelength ranges for the April 1999 and March 2000 runs were
$\lambda \lambda 3400 - 7600$\,\AA\, and  $\lambda \lambda 3900 - 6700$\,\AA.
In both runs we used the 3\arcsec\, diameter Hydra blue fibers.
The field of view of Hydra is approximately 1\deg.

We observed a total of three fields over the course of three nights in 
1999 and the same number of fields in 2000.  Radial velocities 
were obtained in three different areas of the cluster; roughly the central 
part (near M87) as well as the northern ($\delta = 13$\deg: near M86) and 
southern clumps ($\delta = 7$\deg: near M49).  Sample selection was done 
using digitized 
Palomar sky survey plates, in
conjunction with morphological classifications from the Binggeli, Sandage
\& Tammann (1985) Virgo Cluster Catalog (hereafter VCC).  
Each fiber setup contained between 15 to 25 galaxies, with most of the
remaining fibers positioned on the sky across each field.  Some Hydra 
fibers which could not be assigned to dEs were placed on non-dwarf Virgo 
galaxies whose redshifts were unknown.  Total
exposure time for each field was typically about 11,000 seconds.    We bias
subtracted, flat-fielded and then wavelength-calibrated the spectra with a
Helium-Neon-Argon lamp.  All reduction procedures, except the
basic CCD processing, occurred using the {\it IRAF} task {\it DOHYDRA}.
We then combined each of the individual frames and 
later sky-subtracted galaxy fibers using the closest sky fibers to each 
object. Sky subtraction was typically 
successful with an effective removal of the sky continuum and telluric
emission features, with occasional residual sky lines in the red part of the 
spectrum.  These remaining faint sky lines, typically in the 
region near 7000\,\AA\,, were removed by hand.

Long slit data from the KPNO 4m telescope were used to determine the 
radial velocities of two dwarf ellipticals in June 1999.  The RC spectrograph
with the KP-007 grating was used giving a dispersion of 1.4\,\AA\,\
pixel$^{-1}$, and a wavelength coverage $\lambda \lambda$ 4400 - 7350\,\AA.
These spectra were reduced and extracted using the {\it APALL} program in
the {\it IRAF} long-slit package.

During these observations we obtained spectroscopy of
velocity standards to calibrate the velocity zero point.   
The stars observed are listed in Table 1; most are G-K
spectral types.  Each
stellar spectrum was reduced and calibrated the same way as were the galaxy
spectra.  For each night we create
a single velocity template by combining the spectrum of each star observed, 
using
the task {\it SUMSPEC} in the {\it RVSAO} (Radial Velocity Smithsonian 
Astrophysical Observatory) package.

 Using these standards we then computed the radial velocities for our 
complete sample of
observed objects. Only a fraction, $\sim$ 25\%, of the observed galaxies had 
high enough
signal-to-noise (S/N) ratios for a reliable radial velocity measurement.  
Figure 1 shows a typical 
Hydra dE spectrum. We determine velocities by cross-correlating absorption 
features with the velocity template using the {\it XCSAO} task in the {\it 
RVSAO} 2.0 package within the {\it IRAF}
environment (see Kurtz \& Douglas 1998). {\it XCSAO} outputs a radial
velocity for each object computed
from cross correlation fits, including a quality R-value (Tonry \& Davis 1979)
where a higher value of R indicates a better fit, and we adopt R $\geq$ 3 as 
the limit of reliability (Kurtz \& Douglass 1998).  Of those observed, 
43 have high enough S/N and R values $>3$ for a reliable
radial velocity measurement, and 37 of these are dwarf 
ellipticals or dwarf lenticulars.  In
Figure 2, we plot the difference between our computed velocities with
published radial velocities for those few galaxies with previously
measured values.   The average absolute difference between our measured 
velocities
and those previously published is 128$\pm$71 \kms. All new velocities are shown
in Table 2, while Table 3 lists all Virgo dEs with known radial velocities.

We combine our new velocity measurements with previously published ones.  
The most recent previous studies on the 
dynamical properties of the Virgo cluster, Binggeli et al. (1993) and SBB99 
contained 385 and 403 velocities respectively. The present study uses 497 
velocities,  with new velocities mainly of faint galaxies, taken from a 
variety of sources including Drinkwater et al. (1996) and Grogin, Geller 
\& Huchra (1998).   The distribution across Virgo of all the galaxies used 
in our analysis is shown in Figure 3.

\subsection{Galaxy Morphology and Populations}

Identifying a morphological type for a galaxy, or for
our purposes, identifying members of distinct populations, is never a simple
or straightforward procedure.  In Virgo the problem is even greater due to 
the poor correlation of Hubble types with global properties such as star 
formation rates (Koopmann \& Kenney 1998).  A morphological classification
therefore has its limitations.  None-the-less, we
characterize 7 distinct galaxy populations for this study based on
morphology alone.  These  are: classical ellipticals (E), lenticulars (S0), 
spirals (Sa-Sd), 
irregulars (Irr, I, dIrr, Sm, blue compact dwarfs), dwarf ellipticals 
(dE, dSph), nucleated dwarf ellipticals (dE, N), and dwarf lenticulars (dS0).
We often combine dwarf ellipticals and dwarf nucleated ellipticals into
one class which we denote as the total dwarf ellipticals (dET).

Almost all of the morphological types are taken from
the Virgo Cluster Catalog (VCC).  Some galaxies in this paper are 
unclassified in the VCC catalog, particularly in the region 
12$^{h}$50$^{m}$00$^{s}$ - 13$^{h}$15$^{m}$00$^{s}$.  We either classified 
these galaxies ourselves, using second generation
Palomar Sky Survey images, or adopted classifications from other
published studies.  If there were a 
significant  uncertainty in a galaxy's morphological type,  we assigned
no classification and the galaxy was left out of the analysis involving 
galaxy populations.  
The morphological identifications of Virgo core galaxies (defined in the
next section) with radial velocities breaks down as follows: 40 ellipticals, 
56 S0s, 18 dEs, 79 nucleated dEs, 28 dS0s, 81 irregulars and 119 spirals.

\section{Results}

\subsection{Virgo Cluster: Structure}

Smith (1936) was the first to study the dynamical structure of the Virgo 
cluster based on 32 radial velocities obtained by himself, V.M. Slipher
and M.L. Humason (unpublished).  Using this data, Smith (1936) concluded that 
the Virgo cluster is a bound system, and derived a mass for it using the 
kinematics of these galaxies.  This provided some of the first
evidence for extensive dark matter in clusters, similar to the pioneering
study of Coma done by Zwicky (1933).    This derived mass assumed that the
cluster components are virialized; however the Virgo cluster is
not as indicated by the presence of sub-clusters which have been known 
since the study of de
Vaucouleurs (1961).   More detailed observations revealed Virgo to be
a complex system of different aggregate clouds of galaxies
that are part of the Local Supercluster (e.g., de Vaucouleurs 1961; Holmberg
1961; de Vaucouleurs \& de Vaucouleurs 1973; Huchra 1985; Binggeli et al. 1987;
Binggeli et al. 1993; SBB99).     De Vaucouleurs \& de Vaucouleurs (1973) 
identify and named several of these clouds, in
the area of the Virgo cluster studied here ($\alpha$ (J2000): 
12$^{h}$ - 13$^{h}$, $\delta$ (J2000):
 0\deg - 20\deg; v $<$ 3000 \kms).    These are, using the de Vaucouleurs \& 
de Vaucouleurs (1973) notation, the M cloud, Wa and Wb clouds and the Virgo 
I and II clusters.   The W cloud (combined Wa and Wb) is the most distinct,
with a well defined locus at $\delta \sim 5$\deg and velocities
near 2600 \kms. The Virgo II cluster (the B cluster in Binggeli et al. 1987) 
is south of the area used in this study, but it and the other galaxy clouds 
near Virgo are  probably merging with each other based on their relative 
positions and velocities.

These various clouds can bias results; thus to study the Virgo core
cloud we define two separate regions.  We define the ``Virgo
Region'' as those galaxies with $\alpha$ (J2000): 12$^{h}$ - 13$^{h}$, 
$\delta$ (J2000): 0\deg - 20\deg and with velocities v $<$ 3000 \kms.  The 
distribution of 
velocities for each of our defined galaxy populations is
plotted in Figure 4.  This is still a rather generous spatial and velocity
coverage that is certainly
unsatisfactory for studying only galaxies associated with the primary
Virgo core region.  To study only galaxies 
in the cluster proper we identify a separate ``Virgo Core''
containing  only galaxies within 6\deg of the center of the
luminosity density of the Virgo I or A main cluster (defined by Binggeli
et al. (1987) as (B1950) 12$^{h}$ 27$^{m}$ 50$^{s}$; +13\deg 01\arcmin
24\arcsec) and with 
radial velocities v$<2400$ \kms.  Spatially these galaxies are found within
the circle shown in Figure 3.  Isolation of this sample 
allows us to perform statistical tests without
a large bias, or contamination from Local Supercluster clouds seen in
projection towards the cluster, although some contamination is inevitable.
The ``Virgo Core'' velocity histograms are plotted in Figure 5; while the
distribution of spatial positions of all likely members of Virgo, 
based on VCC estimates, are shown for each galaxy population in Figure~6.
Figure 6 contains an essentially complete listing of all Virgo galaxies down to
magnitude B = 18.

The analysis presented here is only concerned with these 
two areas in Virgo, but mostly with the Virgo Core region. While our 
large sample of 497 radial velocities,
429 of which are in the Virgo Core, permits a detailed dynamical study of
the cluster and its neighboring clouds, we postpone an analysis of these
clouds for another paper.  It is sufficient for comparison
purposes to state a few basic parameters (Table 4).  The mean 
Virgo Core heliocentric velocity is found to be 1064$\pm$34 km s$^{-1}$, 
less than 0.5$\sigma$ from the value of 1050$\pm$35 \kms found by 
Binggeli et al. (1993) and has a total velocity dispersion of 705 \kms, close
to the 699 \kms dispersion found by Binggeli et al. (1993).

The usefulness of these mean and dispersion quantities is limited since as 
discussed above, the Virgo 
cluster is not a simple relaxed system with a Gaussian velocity
distribution (see Figures 4-8).    Figures 7 shows the velocity
of each Virgo galaxy relative to the mean Virgo velocity of 1064 \kms, as a 
function of the absolute magnitude. The dashed lines show the escape velocity 
from the Virgo core computed from the mass profile of Giraud (1999).
From Figure 7 we see no equipartition of energy, nor any obvious
dynamical friction effects that would produce lower velocities
for higher mass objects.

\subsubsection{Virgo Clouds: Galaxies Bound to the Cluster?}

Figure 3 shows the velocities of objects across the Virgo cluster.
Triangles represent galaxies with velocities less than 1000 \kms, with 
larger symbols signifying
larger velocity deviations from 1000 \kms.  Circles are galaxies with 
velocities larger than 1000 \kms, with the largest circles signifying galaxies
at v $> 2000$ \kms.
Signatures of various `sub-clusters' or clouds can be seen from the different 
symbols.  The spatial clustering in Figure 3 is however partially
the result of selection 
biases introduced by spectroscopic sample selection.  

There is evidence for spatial clustering in Virgo, as is shown in Figure 6. 
SBB99 found that galaxy clustering 
correlates with the intensity of X-ray emitting gas in Virgo.  In the north, 
there is a sub-cluster with lower velocities associated with the
Virgo cloud around M86.  Likewise, we see velocity structures near 
M87 (the X in Figure 3) and the W cloud near (J2000) ($\alpha, \delta) 
\approx$ (12.35, 8.00).

The dEs also appear to clump slightly around the three major
elliptical galaxies in Virgo: M87, M49 and M86.  Within a degree of each of
these galaxies the velocity dispersion of dETs is essentially the same as the
dispersion of all Virgo dETs.
For the five dETs within 0.5\deg of M87, the dispersion is 428 \kms with a 14\%
change of being random, and
an average velocity of 1097 \kms.  The velocity of M87 is 1307 \kms,
and thus there appears to be a strong clustering of a few
dEs around it.   The 18 dEs within 0.5\deg of M86 (with v = -244 \kms) have 
an average velocity of 508$\pm812$ \kms, which has only a 0.1\% change of 
being 
random.  M49, located at the southern
part of the Virgo cluster, is spatially a separate system, but has a similar
velocity (997 \kms) to that of the Virgo core region.  We see some
evidence for dET association with this elliptical galaxy and its surrounding
cloud.   Within 0.5\deg of M49 there are no dETs, but within 2\deg there are 
six dETs with an velocity of 1083 \kms and a low
velocity dispersion of 291 \kms, with a 2\% change of being random. This 
dispersion remains low (473 \kms)
for all dETs within 3\deg of M49.  This has been known for
some time for all galaxies in the M49 region (Binggeli et al. 1987).
We conclude, in agreement with
Ferguson (1992), that there is some evidence for bound dE companions to giant
galaxies in Virgo, but it is only a statistically significant effect 
for M86.  As is suggested 
from their spatial 
positions, most dEs are distributed in the general gravitational 
potential of Virgo and are not companions of larger galaxies.

To determine if the velocity and projected locations of the various galaxy 
populations are statistically distinct, we perform Kolmogorov-Smirnov (K-S) 
tests on the positions and velocities of each individual core
population against each of the other individual core populations.  We do
not perform these tests for the total galaxy populations, but only the
core ones, to avoid biases from other 
Virgo galaxy clouds. These are two sided K-S tests that compare the 
distributions of velocities and positions of galaxies in
different populations 
with each other (e.g., Press et al. 1992).  Since K-S tests are based on 
statistics of two  vectors, the positional test is limited to the 
projected radial distance from the center of the cluster.   These tests
are also somewhat constrained since if galaxies are in equilibrium
their spatial positions and velocities should reflect the underlying mass 
distribution
of the cluster.  As such the interpretation of a lack of a correlation
between two populations is not straight forward. On the
other hand if two populations were accreted into Virgo at the same time,
then we might expect their orbital and spatial positions to be correlated.

The results of these tests are shown in Tables 5 and 6.
Unfortunately, the results are largely inconclusive showing no statistically 
significant correlation between any two populations, but they are 
instructive in several cases.   We discuss individual K-S test for each
Virgo galaxy population in the next section.

\subsection{Accretion and Orbit Signatures}

Figure 9 shows pie diagrams of slices through the total Virgo sample in both 
right ascension and declination.  
These figures have several features that illustrate the dynamical
state of the Virgo cluster.  The first is the clear `finger of god' signature
of the virialized Virgo core.  The population here is
a mixture of all morphological types.

Also in these figures are clear indications of galaxies still in the
process of accreting 
into the Virgo core.  These produce shell-like `infall ring' regions,
perpendicular to our line of sight in a pie diagram
(Praton \& Schneider 1994).   In Figure 9 there are also several clumpy
areas that contain galaxies, sometimes very close to the `finger of
god' feature.  By far the most common galaxy to deviate from the `finger'
are the spirals which define these clumpy regions. 
The dETs (open circles) tend to follow both the spirals (X figures) and 
ellipticals (black squares);  outer associations are dominated by spirals
and dwarf ellipticals, but with a very high spiral/dET ratio.

While velocities of spirals and  ellipticals in Virgo are almost all known, 
only a very small fraction of dET velocities have been measured.  Of
these,  most are for dETs in
the densest regions, and mainly for the more easily observed dE,N types.
It is therefore possible that dETs populate all 
regions of this diagram, especially in the outer associations dominated
by later-type galaxies.   
Furthermore, velocity dispersions of various Virgo galaxies as a 
function of distance from the cluster core (Figure 10) indicates that dEs
are on complex orbits.  This is also true for 
the spirals, S0s,
dS0s and irregulars, indicating that these systems have a mix of orbits 
including highly elliptical or radial ones.   

\subsection{Velocity Distributions}

If some galaxies in Virgo originate from an accreted component that 
underwent interactions with other cluster members, and if the velocities
reflect only the gravitational potential well of Virgo at time of infall, 
then the various velocity dispersions of accreted populations should be 
somewhere between that of the 
old giant E component, assuming they trace the early
potential (see \S 3.3.1), and currently infalling spiral and 
irregular galaxies.  The evolution of the distribution of 
velocities of Virgo galaxy populations and their dispersions can be
predicted by using reasonable assumptions about their total 
lifetimes in the cluster.  This can be done by investigating the effects of
two-body relaxation, equipartition of energy, and dynamical friction.

To test these ideas we produced velocity histograms (Figures 4 \& 5)
and statistical data (Tables 5 \& 6) on
the total (Figure 4) and core (Figure 5)  Virgo galaxy populations, including:
giant ellipticals (E), spirals (Sa-Sd),
dwarf ellipticals (dE) nucleated dwarf ellipticals (dE,N), lenticulars
(S0), dwarf lenticulars (dS0), and irregulars (dIrr, I, Sm, BCD).    

We fit Gaussian distributions to all of the core histograms (Figure 5)
by using the Levenberg-Marquardt (LM) algorithm  (Marquardt 1963).  We use
the results of these fits to test for kinematic characteristics of infalling 
versus bound galaxy 
populations. In particular, a Gaussian form is expected for a virialized
population.  Thus finding how well a Gaussian can fit a velocity distribution
is one method for determining if a population might be relaxed. 
The Levenberg-Marquardt method uses a non-linear least squares routine to 
fit the number of galaxies (N) as a function of velocity (v) in the form:

\begin{equation}
N = A + B\times {\rm v} + \Sigma \left( C \times {\rm exp}{\left[-2.7\times
\left(\frac{({\rm v}-{\rm center})}{{\rm FWHM}}\right)^{2}\right]}\right),
\end{equation}

\noindent where $A$ and $B$ are constants, and always fit to zero, $C$ is the 
amplitude of the Gaussian,
while ``center'' and ``FWHM'' are the center and full width at half maximum 
for the fitted Gaussians.  This summation is over the total
number of Gaussians needed to fit the distributions.   For cases where either
one or two Gaussian components could be fit, we present in Figure 5 only 
single Gaussians. The centers of the peaks are also fitted by the 
LM algorithm, and not chosen a priori by us.

The different velocity distributions for galaxy populations in the 
total Virgo sample can be seen in Figure 4.  
The ellipticals, spirals,
lenticulars and irregulars have bimodal distributions.  The second 
peak located near 2500 \kms is largely the result of projected clouds that 
are probably falling into the main cluster  (Virgo A in
Binggeli et al. (1987) and Virgo I in de Vaucouleurs \& de Vaucouleurs
(1973)).   Interestingly, there is little evidence of this second 2500 \kms
peak in the dET and dS0 populations.  This is an indication that
nucleated dwarf ellipticals are abundant only in the well defined rich
Virgo cluster proper, located near a velocity of 1000 \kms.  The 
nucleated dwarf ellipticals and 
dS0 population are associated with the dynamical ``Virgo Core'' or Virgo A 
cloud at V = 1064 \kms, although significant sub-structure is evident.

\subsubsection{Ellipticals}

  The radial velocity distribution of ellipticals changes 
slightly between the total 
Virgo area and the Virgo core (Figures 4 and 5), with a 75\% probably
of association from a K-S test. The secondary distribution
centered at 2500 \kms is almost completely absent in the Virgo core 
population, and the radial velocity
distribution of the core ellipticals can be fit as a single Gaussian 
distribution with the best fit $\chi^{2}$ = 1.66, peak at 1093 \kms,
and a velocity dispersion $\sigma = 462$ \kms.

Despite this, even in the core population, there are ellipticals with highly 
deviant velocities for their magnitudes (Figure 8).  One galaxy at 2284 \kms, 
NGC 4168, is uncertain
as a member of the Virgo cluster proper (VCC), and has an uncertain 
classification due to its probable low mass (Ho et al. 1997).  Another
high relative radial velocity galaxy is NGC 4473 with v = 2240 \kms.
It is also doubtful that this galaxy is a proper member of the Virgo Core or
even an elliptical.
It contains disky isophotes, evidence of dust (Michard \& Marchal 1994; van
den Bosch et al. 1994) and its properties do not fit elliptical
galaxy scaling relations (Davies et al. 1983).

Excluding these two galaxies from the elliptical population we see
a slight, but statistically insignificant, signature of equipartition of energy
or effects of dynamical friction. The giant E galaxies appear by eye to have a 
decreasing dispersion with increasing mass, a possible signature of a process 
that has led to the giant ellipticals being more dynamically bound in the 
cluster.  To test if this result is statistically significant, we carry out
a Monte Carlo simulation of the velocities and magnitudes for
these ellipticals.  We characterize the mass-velocity segregation by
measuring the difference between the velocity dispersion for the
bright (M$_{V} < -19$) and faint (M$_{V} > -19$) ellipticals.  The
velocity dispersions of these two components are 299 \kms and 509 km s$^{1}$,
respectively, a 210 \kms difference.  Using Monte
Carlo techniques, we find a 2\% chance that this difference
is random, and that the observed difference is a 2.3$\sigma$ event.  It 
is therefore likely that this magnitude-velocity difference is due to 
chance, and not a real physical effect.  This signature is
removed completely if we include the two rejected ellipticals.

The ellipticals in the Virgo core region have the lowest total 
velocity dispersion $\sigma = 462$ \kms of the major cluster populations, a 
Gaussian velocity distribution 
(Figure 5), and a centrally concentrated spatial distribution (Figure 6). These
all suggest that ellipticals are the oldest and most relaxed component in 
Virgo.   Distances derived from surface brightness fluctuations also suggest
that most of the ellipticals in Virgo are gravitationally bound 
to the cluster (Neilsen \& Tsvetanov 2000). 
Figure 10 shows the steadily decreasing value of the projected
velocity dispersion as a function of projected distance from the dynamical 
center of Virgo for the elliptical population.   Virgo
ellipticals appear to be more dynamically relaxed in comparison to other 
populations.
These effects are also seen for relaxed populations in N-body
simulations of structure formation (e.g., Crone, Evrard \& Richstone
1994; Thomas et al. 1998).  

\subsubsection{Lenticulars (S0s)}

Depending on their nature, cluster lenticulars (S0s) may provide a critical 
clue to the evolution of galaxies in dense extragalactic environments 
(Dressler 1984). The major question to answer is whether or not this 
population has an origin more similar to the 
ellipticals or to the spirals.  Being a `transition-type', the lenticulars 
have properties of both ellipticals (old stellar populations, symmetric 
structure) and spirals (outer disks). 

The kinematics of the S0 population in Virgo clearly differ from those of the 
ellipticals.  The velocity dispersion
of the S0s is 647 km s$^{-1}$, almost 200 \kms higher than that
of the elliptical
population, and similar to that of the dET population.  Based on
their radial velocities and positions, there is a 62\% - 68\% probability
that the dEs originate from the same population as the S0s, although 
there is a lower 5\% - 39\% probability for a velocity association with the 
dE,Ns.   The S0s are
clearly more extended spatially than are the ellipticals (Figure 6) and again
resemble the spirals and dEs in this aspect.  Their 
velocity distribution also contains a low velocity component near 300 \kms
that is also seen in the dET,
spiral and irregular populations.   The overall velocity distribution for S0s
is not fit very well by a Gaussian.  The clumpy distribution of S0
velocities indicates that these galaxies are not a relaxed Virgo component.

Based on this we conclude that S0s are {\it not 
associated} with ellipticals, but formed later, potentially by a  
process similar to that by which the dEs form (cf. Mao \& Mo 1998; 
Moore et al. 1999).  This is consistent with 
observations of higher redshift $z \sim 0.5$ clusters,
where the S0 fraction is as much as three times lower than that
seen in nearby clusters (Dressler et al. 1997), while the fraction of 
ellipticals in clusters is found to be the same at $z<1$ 
(Dressler et al. 1997).   High-resolution N-body simulations of cluster 
formation based on a CDM scenario also show that the properties of
S0s cannot be reproduced from major mergers, while those of elliptical 
galaxies can (Okamoto \& Nagashima 2001), suggesting that S0s formed later.

\subsubsection{Spirals and Irregulars}

The spiral galaxy population has a wide non-Gaussian velocity distribution 
with a total core velocity dispersion $\sigma \sim 776$ \kms, over
300 \kms higher than that of the ellipticals (Table 4). 
The multi-peaked and broad velocity histogram (Figures 4 and 5) can be
interpreted 
as a result of spiral-rich groups infalling into the cluster.
Huchra (1985) characterized the velocity structure of the Virgo spirals 
as `flat-top'.  However our lower errors allow finer binning, and there 
appear to be three components centered near 300 km s$^{-1}$, 1100 \kms and 
1900 km s$^{-1}$.  Similar
peaks are seen in the dET, S0, and irregular populations.  For an
infall model, the high velocity components
are spirals falling into Virgo on the side nearest the Local Group, while
the low velocity components are spirals falling towards us on the opposite
side.  

Further evidence that some Virgo spirals were recently accreted
is shown by a pattern of lessening gas depletion within spirals 
at greater projected radii from the central cluster core, where
the hot dense ICM could removed gas by ram-pressure (e.g., Haynes \&
Giovanelli 1986; Cayatte et al. 1990; Solanes et al. 2001).  Spirals far from 
the core are mainly falling into the cluster, while those in 
the inner part of Virgo have begun to go through the core, and as a result of
gas stripping by the ICM, are depleted of gas. Distances
to Virgo galaxies derived from the Tully-Fisher relation also reveals
the expected infall velocity pattern: galaxies with blue shifts relative
to the cluster are furthest away (Gavazzi et al. 1999).

The spirals in Virgo are statistically similar in velocity space only to the 
irregulars (Table 5).  Is it possible that irregulars were once satellites
of present day Virgo spirals?  During infall companions of spirals will
be separated due to tidal forces. The potential of Virgo will induce, for pure 
linear motion and linear arrangement between, for example,  a spiral of 
mass M$_{{\rm sp}}$ and a companion, a maximum radial differential 
acceleration of

\begin{equation}
{\rm a_{clu}} = {\rm GM_{{\rm clu}}} \left[\frac{1}{{\rm r^{2}}} - \frac{1}{({\rm (r + R)^{2}})} \right].
\end{equation}
 
\noindent The spiral and its companion will become dislodged when
${\rm a_{clu}} >$ GM$_{{\rm sp}}$/R$^{2}$. In these equations, R is the 
distance between the infalling spiral and its companion, r is the distance 
from the leading edge of the group to the center of the cluster, 
M$_{{\rm clu}}$ and 
M$_{{\rm sp}}$ are the masses of the cluster and infalling spiral, and G is
the gravitational constant.  This formula changes somewhat in a non-radial
situation, since eq. (2) is only valid for a radial force.
If we take the mass of Virgo as 10${^{14}}$ 
\solm, concentrated within r, and an infalling spiral galaxy with mass 
10$^{10}$, then a 
galaxy separated by 0.5 Mpc from the host spiral will become dislodged 
within 2.5 Mpc of the cluster center.   We can generalize this by using
radial tides on an infalling galaxy of mass M$_{\rm gal}$ and a companion
that will become
dislodged when RGM$_{\rm clu}$/r$^{3}$ = GM$_{\rm gal}$/R$^{2}$, which occurs
at r = 10 Mpc.

The irregulars\footnote{It is worth repeating for clarification
that our irregular population includes Magellanic Spirals (Sm), as
well as the classical dwarf irregulars and blue compact dwarfs.  Any
comparisons to other studies should keep this in mind.} in, or projected 
against, the core have characteristics very similar to the
spirals and in some ways also to the dwarf ellipticals.  The low and
high velocity components are present in the irregular population (Figure 5) and
the spatial distribution is very similar to that of the spirals (Figure 6).
There is also a 92\% probability that the irregulars
originated from the same velocity population as the spirals, as shown by 
K-S tests (Table 5).  

From the triple peaked velocity distribution and the high velocity
dispersion of 727 km s$^{-1}$, similar to that which we see for the spirals, 
it seems likely that the irregular population
is infalling into the cluster along with the spirals (e.g., Gallagher 
\& Hunter 1989).   Their high gas content and star 
formation rates indicate
these objects have not yet crossed the cluster core, where their extended
gas will be rapidly stripped by the intracluster medium (cf. Lee et al.
2000; Mori \& Burkert 2000). 

\subsubsection{Dwarf Ellipticals}

The overall dET population has a velocity distribution that is similar to 
that of the spirals but not to that of the ellipticals, with
a velocity dispersion of 726 \kms, compared to the value of 776 \kms
for the spirals, and 462 \kms for the
ellipticals.   This general trend has been suspected for some time and found
by previous studies (e.g.,
Bothun \& Mould 1988), but also has been disputed on the basis of smaller
samples than
the one used here (Binggeli, Tammann \& Sandage 1987). 
The projected spatial distribution of the dETs, like that of the spirals, does 
not show as significant a decline in radius as does that of the ellipticals.  
To which 
of these two major populations are the dETs similar?   The velocity 
distributions and dispersions of both dETs and pure dEs (shaded histograms 
in Figures 4 \& 5) resemble more the spirals rather than the ellipticals.

The distribution of dET velocities is not fit by a single Gaussian, 
another indication of dynamical evolution, youth and/or substructure. 
If Virgo dEs were in equipartition 
with the giant ellipticals we would see evidence for mass-position
segregation. In this case, the dEs would have a more extended spatial
distribution, and a larger velocity dispersion than what is observed
(\S 4.1.1).  In 
fact, many dEs are in the richest part
of Virgo, with fewer found in the lower density areas (Figure 6). Possibly
some mass-position segregation is occurring, since the dEs are more
spread out than the giant Es, but this is only a slight effect that could
also be accounted for by dEs originating from accreted components not
on radial orbits. The
velocity distribution of the dEs is also clumpy, including evidence for
a low-velocity peak (Figure 5). 
From these data, we conclude that the dETs taken as a single
population are less dynamically evolved than the Virgo cluster core E 
populations, and are not relaxed.   To further test this will require knowing
the 3-dimensional spatial distributions and velocity dispersion profiles
of Virgo galaxies.

\section{Dynamical Models}

Modeling the dynamical evolution of galaxies in Virgo
is critical for understanding these various kinematic characteristics and how
they evolve. In this section, we investigate various ways the velocity
distributions of galaxy populations can change over time. This includes
determining what the velocity distribution would be if all galaxies
have been in the cluster since its initial formation. Based on these 
calculations we conclude, consistent with our observational data, that the 
elliptical population is the oldest, and only relaxed sub-component of the 
cluster, while all the remaining populations have characteristics of later
infall.

\subsection{Virialization and Relaxation}

The time for a galaxy in a  cluster to relax can be approximated
to an order of magnitude in several ways.   One way is to use the two-body 
relaxation time (Chandrasekhar 1942; Spitzer \& Hart 1971; Spitzer 1987). For
a self-gravitating population of identical mass bodies, the two-body
relaxation time t$_{r}$, defined as the time for a galaxy's orbit to become 
deflected by 1 radian, is given by

\begin{equation}
{\rm t}_r = \sigma^{3}/(4 \pi G^{2} {\rm m}_g^{2} {\rm n}_g{\rm ln} \Lambda) \approx \frac{0.06 {\rm N}}{{\rm ln}(0.15{\rm N})} \times {\rm t}_d,
\end{equation}

\noindent where $\sigma$ is the velocity of the galaxy, m$_g$ is the
mass of the galaxy, n$_g$ is the number density of cluster galaxies,
$\Lambda$ is the ratio of max to min impact parameters, N is the number
of particles in the system, and t$_d$ is the cluster crossing time.

To compute the total relaxation time for the Virgo elliptical
population we assume that early in Virgo's history only the ellipticals 
existed.
Assuming a primordial average elliptical mass of $10^{12}$ \solm, an
initial number density $\sim$ 20 galaxy/Mpc$^{3}$, and
a velocity dispersion of $\sim$ 460 \kms (Table 2) gives a relaxation time 
of $\sim 1.5$ Gyr.  The time for a system to completely relax is about 
10 $\times$ 
t$_{r}$, thus it would take $\sim$ 15 Gyr for the elliptical population to 
complete this process.   This is approximately a Hubble time, and thus 
it is unlikely that relaxation has completely occurred.

The relaxation of the Es is however consistent with observations of galaxy 
populations in ($z \sim 0.5$) clusters where the `core' ellipticals seem to be
well in place (Dressler et al. 1997).
Furthermore, the star formation history of moderate redshift clusters suggest 
that the stellar populations of ellipticals are mostly formed by $z \sim 3$ 
(Ellis et al. 1997).  This formation must occur early to account for
the mass-metallicity relationship found in nearby cluster ellipticals 
(Kauffmann 1996).

This also agrees with results from numerical Cold Dark Matter models
which show the first objects to relax in clusters are large
elliptical galaxies (Diaferio et al. 2000). Simulations of cluster formation 
also indicate that the first galaxies formed in clusters should now be 
preferentially at the 
center (White \& Springel 2000; Governato et al. 2001).  The ellipticals are
consistent with being these objects since they are the most centrally 
concentrated population in Virgo.

For a dET population, similar to the present day density of
several thousand dETs in a volume of a few Mpc$^{3}$, the time for 
relaxation is $> 10^{13}$ yrs, assuming all the Virgo mass then was 
contained in dEs. Thus relaxation is not likely to have occurred for any dETs, 
old or young.

\subsubsection{Energy Equipartition}

We can crudely estimate features of dEs if they have achieved energy 
equipartition 
with the giant ellipticals.  In this scenario the average energies
of the elliptical and dwarf ellipticals should be the same.
This would require that the average velocity of the dET population be
(M$_E$/M$_{dE}$)$^{1/2}$ $\times$ V$_E$, where all quantities are averages.
Taking the average mass of a Virgo elliptical and dwarf elliptical as 
10$^{12}$ \solm, and 10$^{8}$ \solm respectively, and the observed velocity 
dispersion of 462 \kms for the ellipticals, the average velocity of
the dEs should be $\sim 46 \times 10^{3}$ \kms.  This velocity is much
larger than the observed 726 \kms and is larger than the Virgo escape velocity.
Dwarfs therefore would be escaping the cluster, and at the very least would 
have a very large velocity dispersion.
We should also begin to see some mass segregation,
with significant numbers of dETs in the outer parts of the cluster.   Most 
dETs 
are however in the core region of Virgo, although 
with a large dispersion in distances (Young \& Currie 1995; although see
Binggeli \& Jerjen 1998). 

The time scale for energy equipartition to occur is given by (e.g., 
Bertin 2000, eq. 7.10),

\begin{equation}
{\rm t_{eq}} = \frac{{\rm m_{f}}}{{\rm m_{g}}} \times t_{r},
\end{equation}

\noindent where t$_{r}$ is the relaxation time (eq. 3) and m$_{{\rm f}}$
is the mass of field particles.  For the giant Es, ${\rm t_{eq}}$
$\approx$ $t_{r}$ = 1.5 Gyr, although for dwarf ellipticals to
reach equipartition with ellipticals ${\rm t_{eq}} > 10^{3} \times 
{\rm t_{r}} \sim 10^{7}$~Gyr.   Based on this approximation the dETs and
the other Virgo populations cannot be in equipartition of energy with the 
ellipticals.  This is verified by our observations (Figure 8) that
shows no signs of equipartition in the dET or spiral population.  However, the
giant Es can be in equipartition with each other, if we assume that the
Es were the first and only galaxies in the cluster several Gyrs ago,
and all additional galaxies were accreted adiabatically.

Any new dETs introduced into the cluster, as well as any old dETs, should show 
no signs of energy equipartition or full relaxation. Lacking these signatures 
therefore does not prove that Virgo dEs are a young cluster population.  

\subsection{The Evolution of Velocity Structures and Evidence for Infall}

To understand how any new galaxy population introduced into Virgo will 
kinematically evolve, we consider the case
of galaxies undergoing spherical accretion into a cluster at the turn-around
radius, where the velocity of infall
just cancels the expansion rate of the universe (Peebles 1980).

In the special case where the density perturbation associated with a
galaxy cluster has
a constant density within the turnaround radius R$_{\rm T}$, we can
write the mass within any radius r, where r$<$ R$_{\rm T}$ as M($<$r) = 
(4/3)$\pi$ r$^{3} \rho$.    The infall velocity of a galaxy
joining a cluster with mass M, and turnaround radius R$_{\rm T}$
scales as v$^{2} \sim$ $\rho \times$ R$_{0}^{2} \sim$ M/R$_{\rm T}$.  Thus,
depending on the form of M/R$_{\rm T}$, clusters
with higher masses will contain accreted galaxies with higher radial
velocities.  This
can generalized in the following way (e.g., Gott \& Rees 1975).   When
the initial density enhancements in the universe can be represented by
a power law after recombination, then

\begin{equation}
\frac{\delta \rho(t_{0})}{\rho (t_{0})} = \left(\frac{{\rm M}}{{\rm M_{0}}} \right)^{-1/2 - n/6},
\end{equation}

\noindent where n $\geq -3$ for hierarchical 
clustering.   For a universe with $\Omega = 1$, the mass and size, 
R$_{\rm T}$, of assembled areas are related by

\begin{equation}
\frac{{\rm R_{T}}}{{\rm R_{0}}} = \left(\frac{{\rm M}}{{\rm M_{0}}} \right)^{5/6 + n/6}.
\end{equation}

\noindent Unless M/R$_{\rm T}$ remains static, or decreases, which
requires 
n$\leq$1, then galaxies accreted at later times when the cluster
mass is higher, will have higher M/R$_{\rm T}$ ratios, and thus
higher observed velocities than when the cluster contained less mass.
This is consistent with spirals and irregulars being the most recently
accreted component since they have the highest velocity dispersions.
The rate of shell infall increases with
higher $\Omega_{M}$, and for $\Omega_{M} = 1$ {\it all} shells eventually fall
into the cluster.   The current Virgo mass within a 
radius of 800 kpc is 1.8 - 3.3 $\times 10^{14}$ \solm, based on fits to 
mass profiles (Girard 1999) while SBB99 find a somewhat lower value.
In the past Virgo likely contained a lower total mass, and any galaxies
accreted then will have lower observed velocities than those captured
recently.

Dynamical friction is also an important aspect to consider.  We can
compute the effective time scale for dynamical friction, in terms of
the initial radius $r_{i}$, velocity V and mass M, assuming
a static potential, as (Binney \& Tremaine 1987):

\begin{equation}
t_{{\rm fr}} = \frac{264~{\rm Gyrs}}{{\rm ln} \Lambda} \left(\frac{r_{i}}{2 {\rm kpc}}\right)^{2} \left(\frac{{\rm V}}{250 {\rm~km~s^{-1}}}\right) \left(\frac{10^{6} {\rm M_{\odot}}}{{\rm M}}\right).
\end{equation}

\noindent  For the ellipticals, t$_{{\rm fr}}$ $\sim$ 3 Gyr.  This
is short enough for dynamical friction to have an effect
on the kinematic structure.  This is
especially true if ellipticals have existed in clusters for many Gyrs,
as is suggested by observations (e.g., Ellis et al. 1997) and
theory (Kauffmann 1996).  For all other lower mass
populations, including the dEs, dynamical friction is not an efficient process
with t$_{{\rm fr}}$ $\sim$ 10$^{3}$ - 10$^{4}$ Gyrs.

We conclude from this argument, and the previous calculations in \S 4.1,
that velocities of
low mass ($< 10^{10}$ \solm) accreted galaxies, and hence their velocity 
dispersions, should not change
very much over a Hubble time due to dynamical friction, relaxation or
energy equipartition. If dEs are a new population that resulted from infall 
in the last several Gyr, then they should retain many of their original orbital
characteristics.  Two possibilities then exist for explaining a separation
between giant ellipticals and the dEs. The ellipticals and dEs could have
formed together and later the distribution of ellipticals modified to become
more centrally concentrated due to dynamical evolution. Alternatively, the
properties of the dEs may reflect later infall into the cluster.

We can use the fact that the ellipticals are relaxed
while the other populations should retain their original orbital
features to argue that most spirals and irregulars
have recently been accreted, while the dET and S0 populations primarily
originated
from galaxies that fell in after the cluster core formed.
For a virialized population the kinetic (T) and gravitational
potential energy (U) should be related by $\bracevert$T$\bracevert$ = 1/2 
$\bracevert$U$\bracevert$, while for an accreted population U + T = 0
and thus $\bracevert$T$\bracevert$ = $\bracevert$U$\bracevert$.  Assuming
that on average the potential of a population that fell into Virgo is equal to
the potential of a virialized population, we can relate the velocity 
dispersions of a virialized and infalling population, assuming isotropic
orbits, as (see also Colless \& 
Dunn 1996) $\sigma_{infall}$ = $\sqrt{2} \times \sigma_{vir}$ = 1.41 
$\times \sigma_{vir}$.  For the Virgo populations the ratios of 
$\sigma_{{\rm pop}} / \sigma_{{\rm E}}$, where pop is any one of 
dEs, S0s, irregulars, 
or spirals, are $\sigma_{{\rm dET}} / \sigma_{{\rm E}}$ = 1.57 or 
1.51 if we include dS0s,  $\sigma_{{\rm S0}} / \sigma_{{\rm E}}$ = 1.4, 
$\sigma_{{\rm Irr}} / \sigma_{{\rm E}}$ = 1.57,  $\sigma_{{\rm Spiral}} / 
\sigma_{{\rm E}}$ = 1.65. These values are all very close to that expected
for the ratio between infalling and virialized populations, and are not 
obviously accounted for by the
dynamical evolution models. We therefore favor the delayed infall model
for the source of cluster dEs. If infall is occurring, orbital 
characteristics should also correlate with galaxy types; for example, 
star forming galaxies should have unique orbital properties
(Mahdavi et al. 1999; Balogh, Navarro, \& Morris 2000), including large
velocity dispersions which we see (Figure 10).

\subsection{High-Speed Interactions}

Although neither dynamical friction nor relaxation induce major effects on
the evolution of kinematics of low-mass cluster galaxy, high-speed or impulse 
interactions are potentially an important method for altering the internal
properties of
infalling galaxies. The interactions between galaxies in Virgo, with a total 
velocity dispersion of $\sim$ 700 \kms, are nearly all at high-speed, at
relative velocities much
higher than individual internal galaxy velocity dispersions. Thus, 
it is  instructive to investigate how these interactions can affect the  
properties of cluster galaxies.  As we show here, high-speed interactions,
while important, are rather inefficient at {\em destroying} low-mass galaxies,
but are likely to significantly {\it modify} their structures.

We can approximate the strengths of high-speed interactions by using the 
impulse approximation (cf. Spitzer 1958; Binney \& Tremaine 1987).
The energy imparted into a galaxy during a non-interpenetrating
impulsive interaction can be computed in the tidal approximation 
by (Spitzer 1958),

\begin{equation}
\Delta {\rm E} = \frac{4 G^{2} {\rm M}_{2}^{2} {\rm M}_{1}}{3{\rm b}^{4}{\rm V}^{2}}{\rm r}^{2},
\end{equation}

\noindent where M$_{1}$ and M$_{2}$ are the masses of two systems
undergoing the impulse, b is the impact parameter, V is the initial
relative velocity of the two systems and r$^{2}$ is the mean square
radius of the perturbed system.  This energy typically goes into
the individual stars that can then be liberated from their
host galaxies (Gallagher \& Ostriker 1972; Aguilar \& White 1985).   

An important question to ask is if this internal energy addition will be enough
to destroy individual galaxies.  Figure 11 shows the ratio of $\Delta$E to 
the internal energy, defined as 1/2~M$_{1}$~$\sigma$$^{2}$, of a galaxy 
interacting with a M = 10$^{12}$ \solm system, as a 
function of the impact parameter, b from eq. (8).
Both lines represent systems with internal velocity dispersions 
$\sigma =$ 50 \kms and relative velocity $V = 1000$ \kms,
with the solid line a system with radius r = 5 kpc (a spiral) and the dashed 
line with radius r = 2 kpc (a dwarf).  The ratio of the increase in
internal energy from an impulsive interaction to the internal binding energy of
a galaxy with velocity dispersion $\sigma$ is given by,

\begin{equation}
\frac{\Delta {\rm E}}{{\rm E}_{internal}} = \frac{8}{3} \frac{G^{2} 
{\rm M_2}^{2}}{{\rm V}^2 \sigma^2}\left(\frac{{\rm r}^2}{{\rm b}^{4}} \right). 
\end{equation}

\noindent Figure 11 shows the relative unimportance of this energy increase 
for destroying galaxies undergoing a single impulse. Typically
only dwarf systems that come within 10 kpc of large ellipticals with mass
$\sim$ 10$^{12}$ \solm will receive enough of an increase in internal energy to
disrupt the system in one encounter.  These distances are rarely, if ever, 
reached in a cluster
of galaxies such as Virgo.  In addition, at this close distance the impulse 
approximation breaks down, and these equations no longer hold.
The impulse energy from high-speed interactions is thus unlikely to destroy
the large number of dwarf ellipticals in Virgo at least on short time
scales. This will, however, potentially
have an important effect on each galaxy's internal evolution (see
\S 5.3).

\section{Discussion}

In the previous section we demonstrated how dETs, along with the Virgo
spirals, irregulars and S0s, all have kinematic signatures of past accretion. 
In this section we discuss several possible origins of the Virgo dETs.  This
includes investigating whether or not dETs could be accreted field
dwarfs, or if they formed from some dynamical process induced by the cluster.

\subsection{Infalling Field and Group Dwarfs}

Could a significant fraction of Virgo dETs originate
from field  or group dEs accreted into Virgo? Is it possible that some or all
the dETs in Virgo were
once part of something like the Local Group (LG) that fell into Virgo,
with the dETs dislodging from the spirals?  The 
answer depends on the number of dwarfs in groups and in
the field, an uncertain quantity
(e.g., Ellis et al. 1996; Trentham 1998). We do however have some limits on 
the number of field dwarfs, and particularly the number of 
dwarf galaxies around spirals, such as those in the Local Group.
The number of dwarf galaxies per giant is very high in clusters, much
higher than in the field (e.g., Binggeli et al. 1990; Secker \& Harris 1996).
If all the spirals in Virgo originated in systems similar to the LG
then in the magnitude limit of the VCC (B$_{T} \leq 18$; M$_B < -13.3$ 
assuming m-M = 31.3) there are only five LG dwarf spheroidals that
could be be detected.  These are NGC 147, NGC 185, NGC 205, the Fornax dSph, 
and
the Sagittarius dSph. We consider M32 a stripped giant elliptical, and not
a dwarf elliptical, for these purposes, although including it as a dE does
not change the following results.  

There are only three bright LG spiral galaxies and they are all detectable
with the same magnitude limits: The Milky Way, M31 and M33.   
This gives a dwarf elliptical to giant spiral ratio of 5/3 = 
1.67. In the VCC there are 1277 members, with 574 more possible 
members, most of which are dwarf ellipticals.  There are, however, only 
slightly over 100 spiral galaxies in Virgo ($\sim$ 160 if we include
S0s), which would correspond to less than 300 
detectable dwarf ellipticals if the dET/gS ratio were the same as in
the LG.   Similarly, if the 40 Virgo cluster ellipticals originated from
mergers of spirals, they would also come from an initial population of
$\sim$100 spirals, and we would again expect to find only a few hundred dE
companions.  The VCC has well over 1000 dEs at these limits, and
thus a simple scenario where the Virgo dEs are old members of galaxy
groups dislodged from their spiral parents under predicts the number
of dEs by over a factor of 3.   There is also no sign of excess dEs 
associated with currently infalling spirals.   Since the ratio of E/S0s to 
dEs is lower than that which see in groups we conclude  
that dETs in clusters cannot be accounted for by accretion of field dEs, or
dwarfs attached to infalling spirals that later become removed.

\subsection{Gas Stripping: dI $\rightarrow$ dE}

  One of the historically more popular methods of creating a dwarf elliptical
is by stripping gas from a star-forming dwarf irregular (Lin \& Faber 1981).
This type of evolution is possible in Virgo since the dIs, along with the
spirals, have an infalling signature and thus it is possible that a large
number of dIs were accreted in the past.
We do know that the large number of dETs cannot be accounted for by the 
relatively 
small number of present day dIs (Gallagher \& Hunter 1989), thus a much
higher past dI accretion rate would be required.   We therefore conclude that 
the vast majority of the dETs in this study, that are by necessity bright, 
did not evolve from irregulars. However, because of the huge population of 
cluster dETs and their large range in sizes
and magnitudes, it seems possible that
some of them, particularly the low mass ones, evolved from irregulars.  
Irregulars have low masses and it is possible that after rapid gas stripping
processes such as impulse encounters would produce a very small and faint 
dwarf elliptical after many Gyr.  It would be very useful to extend searches for an extremely faint 
population of dEs that should exist, as products of the evolution of dIs
that have been accreted into clusters over many epochs along
with their spiral companions (cf. Impey, Bothun \& Malin 1988).  Some of
these have potentially been found in Virgo (Caldwell \& Armandroff 2000).

\subsection{Galaxy Harassment: S $\rightarrow$ dE}

Galaxy harassment (Moore et al. 1996; Moore et al. 1998) is the process
whereby cluster galaxies are stripped of their interstellar medium and 
become dynamically `heated' by high speed interactions with
other cluster galaxies and the cluster's gravitational potential.   
During these encounters the internal potential 
energy of an infalling galaxy increases creating a situation where the galaxy
is no longer in equilibrium.  To re-establish equilibrium the galaxy increases
in size and loses its most energetic stars.   Because of this, the galaxy's 
density decreases over time and after several orbits a spiral can 
morphologically transform
into a dwarf elliptical (Moore et al. 1998).  The lost mass necessarily
becomes part of the intracluster material.

Galaxy harassment can explain a number of observations.
Moderate redshift clusters at $z \sim 0.8$ have a high fraction of
spirals and other star-forming galaxies (e.g., Butcher \& Oemler 1978; 
Butcher \& Oemler 1984; Dressler et al. 1994; Oemler, Dressler \& Butcher 1997)
that are not seen in the same frequency in nearby clusters.   These
blue galaxies typically do not have identifiable merging or interacting
companions, and therefore could be products of high-speed interactions, or 
galaxy interactions with the cluster potential inducing star formation.
Features in nearby clusters,
such as debris arcs (Mobasher \& Trentham 1998; Gregg \& West 1998), and
distorted galaxies without obvious companions (Conselice \& Gallagher 1999),
are consistent with the harassment ideas of cluster galaxy evolution.
The intracluster light in nearby clusters has similar colors to dwarf
ellipticals, suggesting that material similar to dwarfs,
perhaps the remnant material from stripping, is the source of this 
light (e.g., Secker et al. 1997).   Stripping is further confirmed by 
finding red giant
branch stars (Ferguson, Tanvir, \& von Hippel 1998) and planetary nebula 
(e.g., Mendez et al. 1997) in the Virgo cluster intragalactic regions.

The kinematic results presented here also support the 
galaxy harassment scenario for a dET origin. 
As shown in \S 4, the velocity characteristics of low-mass 
($< 10^{10}$~\solm) infalling galaxies
should change little over several Gyrs and thus their velocity and
spatial distributions should reflect the mass profile of 
Virgo at the time of accretion (Carlberg et al. 1997).  
Galaxies accreted by more massive clusters will have higher observed velocities
than those galaxies accreted by a lower mass cluster of the same
size.
If galaxies in Virgo were distributed in a similar way, those accreted early 
will have existed in the cluster longer, and thus
should be more dynamically stripped by high-speed interactions than those 
accreted recently.   In this scenario accreted galaxies
with the lowest masses should have the lowest velocity dispersions.
After the giant ellipticals, the dS0s and dEs have the 
lowest velocity dispersions (Table 4) with a combined $\sigma = 621$ \kms.  
The S0s and dE,Ns, generally
more massive than the dS0s and dEs, have a combined $\sigma \sim 705$ \kms,
over 80 \kms\, higher than that of the lower mass dEs and dS0s.  This 
difference and the difference in spatial distributions is
potentially reflecting the different masses of Virgo when each galaxy type
entered the cluster.  This is
also consistent with the spirals and irregulars infalling now, since
these galaxies have a combined $\sigma \sim 756$ \kms, higher
than any of the other galaxy types.  
  
This, however, is circumstantial, since the cluster mass given by each
population depends upon their spatial and velocity distributions.
This can be investigated in detail by using  the Jeans
equation that relates orbital and spatial distributions of
particles in a spherical system to that system's mass profile M(r) (Binney
and Tremaine 1987; Carlberg et al. 1997),

\begin{equation}
{\rm M(r)} = - \frac{{\rm r} \times v^{2}_{r}}{{\rm G}} \times \left(\frac{{\rm d ln \nu(r)}}{{\rm d ln r}} + \frac{{\rm d ln }v^{2}_{r}}{{\rm d ln r}} + \beta \right),
\end{equation}

\noindent where $\nu$(r) is the radial density profile parameter, 
$v^{2}_{r}$ is the
radial velocity dispersion and $\beta$ is the anisotropy parameter, $\beta =
1 - v^{2}_{\theta}/v^{2}_{r}$.  It is impossible to derive
the value of $\beta$ without knowing the mass profile
of Virgo, or obtain the mass profile with out an assumption about $\beta$.
The luminosity density $\nu$, and total velocity dispersion, 
$v^{2}$ are real space quantities, while we observe the
projected luminosity density $N$ and projected velocity dispersion 
$\sigma^{2}$.  Figure 12 shows the projected spatial distribution of
various populations, and Table 7 list the projected slopes for
both the velocity and spatial distributions.   It is possible to convert
these slopes into three dimensional quantities by using the Abel
transformation (e.g., Binney \& Mamon 1982), although we do not have
enough information to do this.  Additionally, only the dETs 
have a projected spatial distribution slope that can be
measured with certainly out to a few degrees in radius.

Other recent evidence of S $\rightarrow$ dE evolution includes
luminosity profiles of dEs and dE,Ns (Stiavelli et al. 2001), demonstrating
that dETs have inner profiles more similar to spiral bulges than those of giant
ellipticals.  Many Virgo dETs also contain disk-like, as opposed to 
elliptical-like
profiles (Ryden et al. 1999) and there are examples of Virgo dETs 
with faint spiral structure, including IC 3328 (Jerjen, Kalnajs \& Binggeli 
2000). There are also Virgo dETs with dust and gas features commonly associated
with late-type star forming galaxies (Elmegreen et al. 2000).  These 
observations all suggest that dETs could be descendents of star forming
disk galaxies.

How could so many cluster dEs (VCC) originate from spirals?
The rate of infalling galaxies into clusters is predicted to peak
at $z \sim 0.8$ in a CDM cosmology with $\Omega_{M} = 1$, $H_{0} = 50$ \kms
Mpc$^{-1}$ (Kauffmann 1995). This is
the same redshift where the Butcher-Oemler effect is found and where the
infall rates into clusters is observed to be high (Ellingson et al. 2001).  
If dEs originate from spirals, then a significant number of them should 
have been introduced into the cluster at approximately this redshift.  Since 
dEs are
galaxies with infall signatures, including a non-Gaussian velocity 
distribution, that cannot be intrinsic to an old
cluster population, or to accreted field dEs, they likely originate
from this epoch.  They are therefore possibly the modern descendents of 
harassed Butcher-Oemler galaxies.

\subsubsection{dEs vs. dE,Ns}

The velocity 
characteristics of dEs and dE,Ns are not significantly
different (pure dEs are shaded in Figures 4 \& 5).  This suggests that these 
two populations have similar origins and that dE,Ns are {\em not} scaled down 
giant ellipticals.  The one possible difference between dE and dE,Ns is their 
spatial distribution (Figure 6), with the dE,Ns appearing more centrally 
concentrated in Virgo (Ferguson \& Sandage 1989). Analysis of VCC positions 
for the dEs and dE,Ns shows that 
although the average distance from M87 is different by 0.31\deg for the dEs 
(2.25$\pm$1.73\deg) and the dE,Ns (1.94$\pm$1.64\deg),
this difference is not significant when we consider the large $\sim 1.7$\deg
dispersions,  and when errors are taken into account.

Ferguson \& Sandage (1989) found a luminosity dependence in clustering for
Virgo dEs, such that dE,Ns and the faint non-nucleated dEs are 
centrally concentrated, but the bright non-nucleated dEs are not.
There are several possibilities for explaining these trends.  The nuclei of
dE,Ns could be induced from gravitational effects due to
the cluster core.  This can occur in two ways.  One could be a gravitational
effect similar to the one producing the large
fraction of barred spirals found in the Virgo core region (Andersen 1996),
where tides could induce an increase in 
stellar density at the cores of dwarf galaxies. 
On the other hand the gravitational potential could also stabilize dEs in
the central regions of Virgo, allowing globular 
clusters to accrete to the center by dynamical friction,
forming nuclei (Hernandez \& Gilmore 1998; Oh \& Lin 2000; Lotz et al. 2001).
In addition, about 20\% of Virgo dEs have nuclei offset from their 
centers (Binggeli, Barazza \& Jerjen 2000).  This is a possible indication 
that external gravitational fields are driving the nuclei to oscillate about 
galaxy centers, and thus playing a role in dE nuclei evolution.    

Another possibility is that nucleation
is produced by star formation induced in the centers of these galaxies
after they are accreted into the cluster.   Any spirals
that orbit close to the center of the cluster will experience 
gravitational torques which could induce gas to inflow into a dE,N progenitor's
center producing bursts of star formation.  Nucleated dEs also have
a higher specific globular cluster frequency than non-nucleated dEs (Miller
et al. 1998), an observation explainable by increased star formation in  
dE,N progenitors that orbit close to the Virgo core.

These observations suggest that the harassment process 
of transforming spirals into dEs is likely occurring, but
probably is not the only source of these galaxies.  Since most galaxies
are in groups, and these fall into clusters, it seems certain that {\em some} 
dEs in Virgo were formed outside the cluster in galaxy groups.    Our 
observations similarly do not rule out the possibility that some 
dEs in the core of Virgo were present when the cluster Es formed.

\section{Summary and Discussion} 

In this paper we derive the following conclusions based on the kinematic 
properties of Virgo cluster galaxies:

\noindent I. Elliptical galaxies in the Virgo core region form a relaxed,
or nearly relaxed
system.  This is demonstrated by the Gaussian velocity
distribution of the ellipticals, and their centrally concentrated spatial
distribution.  No other galaxy population has characteristics of
relaxation.  This includes the S0s, dEs, dE,Ns, dS0s, as well as the spiral
and irregular galaxies. Relaxation and dynamical friction are shown to be 
inefficient for significantly decreasing the velocity dispersion of galaxies
with masses lower than 10$^{10}$ \solm, although these are potentially 
important for the ellipticals. Infalling galaxies should therefore
retain their kinematic signatures even after many cluster crossings.

\noindent II.  All populations, except for the ellipticals, have similar 
velocity dispersions (Table 4), with an average 
$\sigma_{{\rm pop}}$/$\sigma_{E}$ = 1.49, close to the expected ratio of 
velocity dispersions for a marginally bound and virialized 
population.   The dET, S0, spiral and irregular galaxies have spatial 
distributions that are progressively less centrally concentrated and 
have increasingly non-Gaussian
velocity distributions. We therefore conclude that most galaxy 
populations in Virgo, besides the giant ellipticals, have been accreted 
into the cluster after the giant Es were in place.

\noindent III. We investigate several possible origins for the dEs in Virgo
including: dislodging of dEs companions connected to infalling spiral hosts,
transformation of infalling dIs into dEs, and stripping of material from 
infalling galaxies
through processes such as harassment.  We favor the last scenario with its 
implication that accreted spiral galaxies in the past were transformed into
modern cluster dETs.  
This is suggested in part:  the
least massive galaxies, and those predicted to be the most
stripped (dS0s and dETs) 
have the lowest velocity dispersion (621 \kms) of any population besides the 
large ellipticals.   The spirals and irregulars, the
newest additions to the cluster, have a combined $\sigma$ = 756 \kms, the
highest for any population, potentially reflecting the higher mass of Virgo 
when the dE progenitors were accreted. To confirm this will require 
a better understanding of
the velocity anisotropy of the various galaxy populations and their three
dimensional spatial distributions.

These results have several implications.  First, most dETs in clusters are not
primordial cluster galaxies, or galaxies that have existed since the large
ellipticals were in place.   Since dETs are common in clusters, 
have dynamical properties that suggest they were accreted by Virgo several 
Gyrs ago, and were more massive in the past, they are potentially the remnants
of Butcher-Oemler galaxies.  This raises the question of what happened to 
the original Cold Dark Matter objects?  Did they {\em all} merge early to 
form the giant elliptical population, or is there a large population of 
objects with masses and sizes similar to dEs, but with low surface brightness, 
that remain undetected (Cen 2001), or was the formation of low-mass
halos suppressed (Bullock, Kravtsov \& Weinberg 2001)?  Either solution for 
explaining away the lack of detected old low-mass cluster objects
has deep implications for observational cosmology and deserves further 
attention.

This study suggests that many cluster dEs are fundamentally different
from dwarf ellipticals, or dwarf spheroidals, found in groups, although
they could both be produced by harassment (Mayor et al. 2001).  Establishing
if this is true will 
require more observational evidence to show conclusively, but there may be
two ways to form dEs - a simple galaxy collapse (e.g., Dekel \& Silk 1986),
and stripped down galaxies as suggested here.  This paper also
demonstrates how galaxy evolution can affect cosmological measurements.  
Galaxy clusters would be brighter if harassment did not occur, since
galaxies are stripped of their stellar material.  As a result this can change 
our estimate of cosmological biasing in dense regions.

Future papers in this series will explore the properties of the stellar
populations in cluster dEs themselves to learn more about their histories, 
and to identify exactly
what their ancestors were.  If cluster dEs are ``primordial'' cluster members
then their stars should all have roughly the same age, with no star formation 
after their initial creation, due to ram-pressure effects. If cluster dEs 
form by some other
process, such as harassment of infalling spirals that were once
BO galaxies, then there should be significant intermediate or young stellar
populations. The stellar population ages of dEs should also correlate with 
their orbital kinematics.

We thank
Linda Sparke, Harry Ferguson, Sydney Barnes, Gus Oemler and Elizabeth Praton
for comments and suggestions regarding this work. We also 
thank Bruno Binggeli for the electronic version of the VCC. The referee 
Bryan Miller
and scientific editor Greg Bothun raised several issues that improved
the presentation of this paper.  This research was supported 
in part by the National Science Foundation (NSF) through grants AST-980318
to the University of Wisconsin-Madison and AST-9804706 to Johns Hopkins
University. CJC acknowledges support from a Grant-In-Aid of Research 
from Sigma Xi and the National Academy of Sciences (NAS) as well as a 
Graduate Student Researchers Program (GSRP) Fellowship from NASA and a 
Graduate Student Fellowship from the Space Telescope Science Institute (STScI).

\clearpage

\input tables_input.tex

\clearpage

\input figures_input.tex

\end{document}

%% file: tables_input.tex
\begin{figure}
\plotfiddle{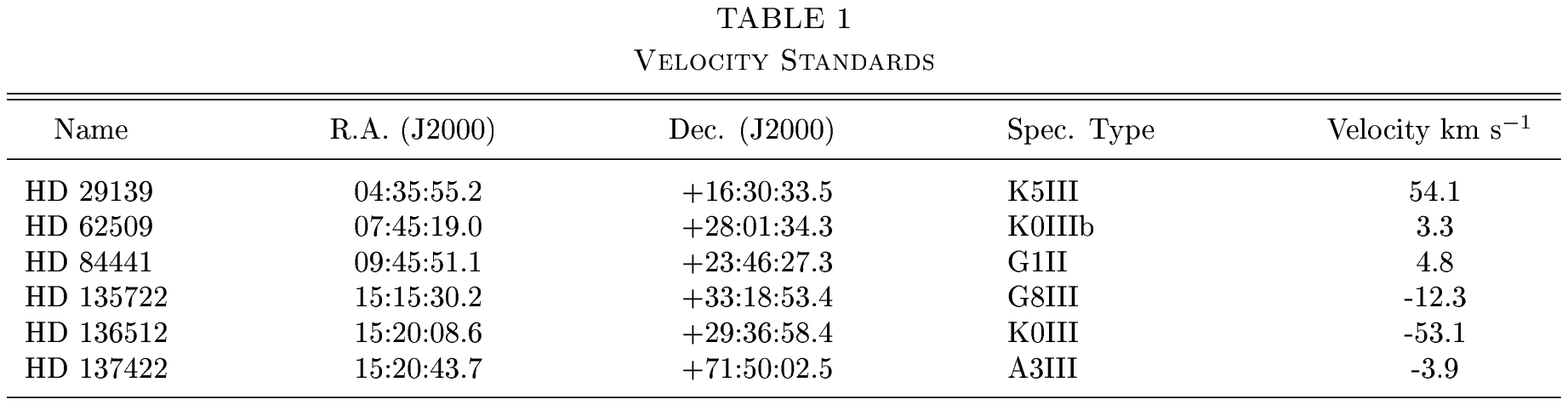}{6.0in}{0}{100}{100}{-310}{-170}
\end{figure}

\begin{figure}
\plotfiddle{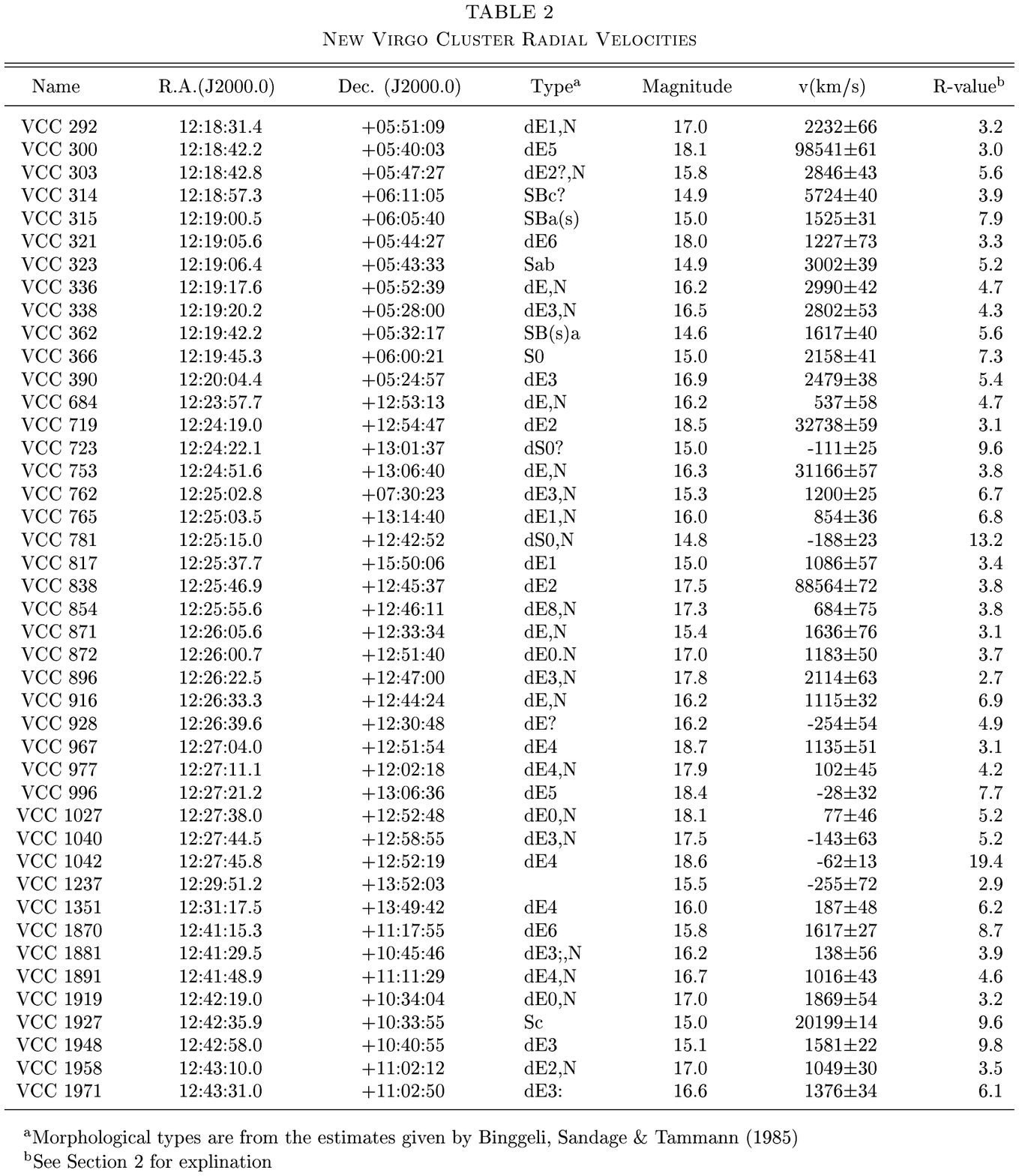}{6.0in}{0}{100}{100}{-310}{-170}
\end{figure}

\begin{figure}
\plotfiddle{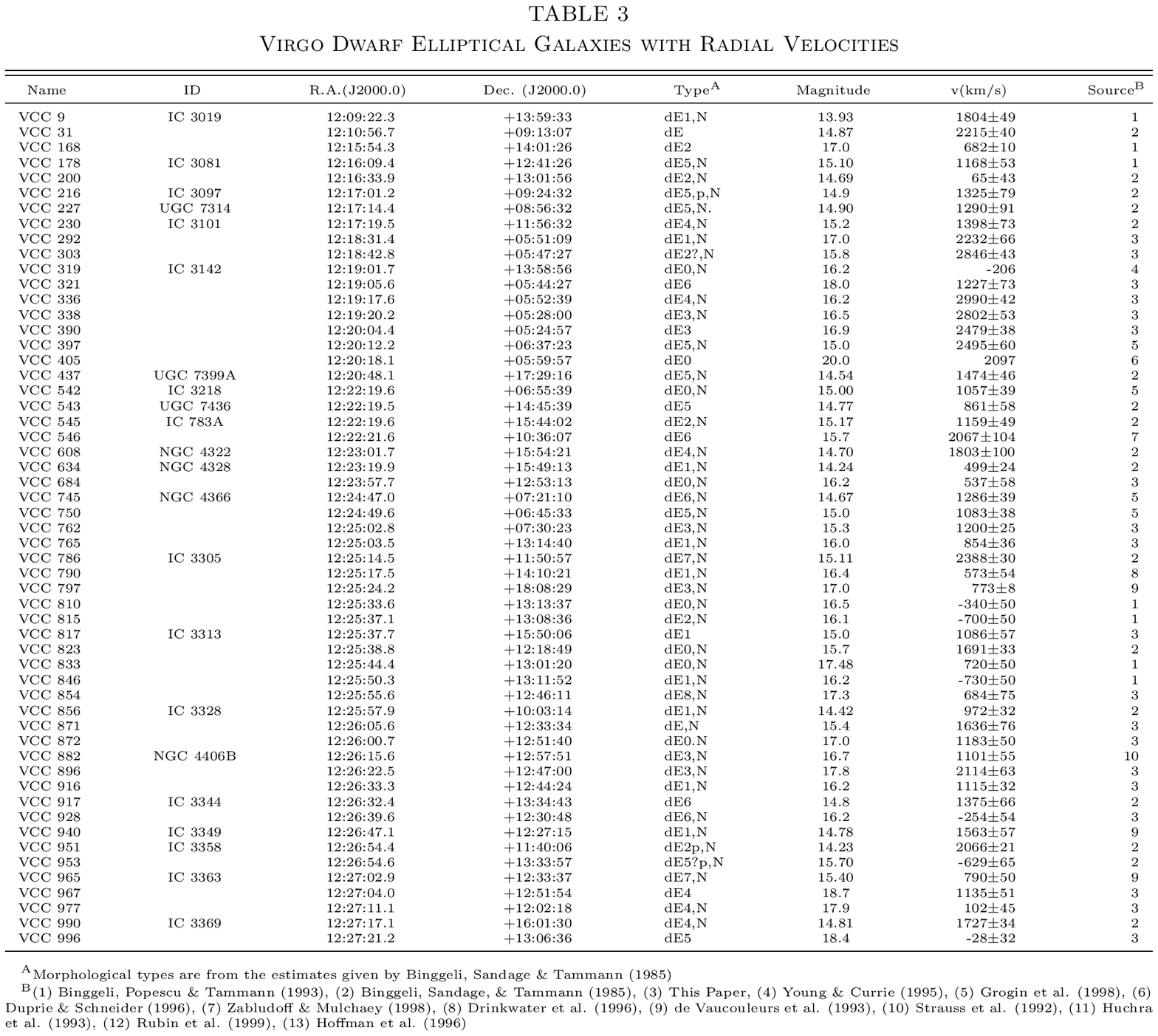}{6.0in}{0}{100}{100}{-310}{-170}
\end{figure}

\begin{figure}
\plotfiddle{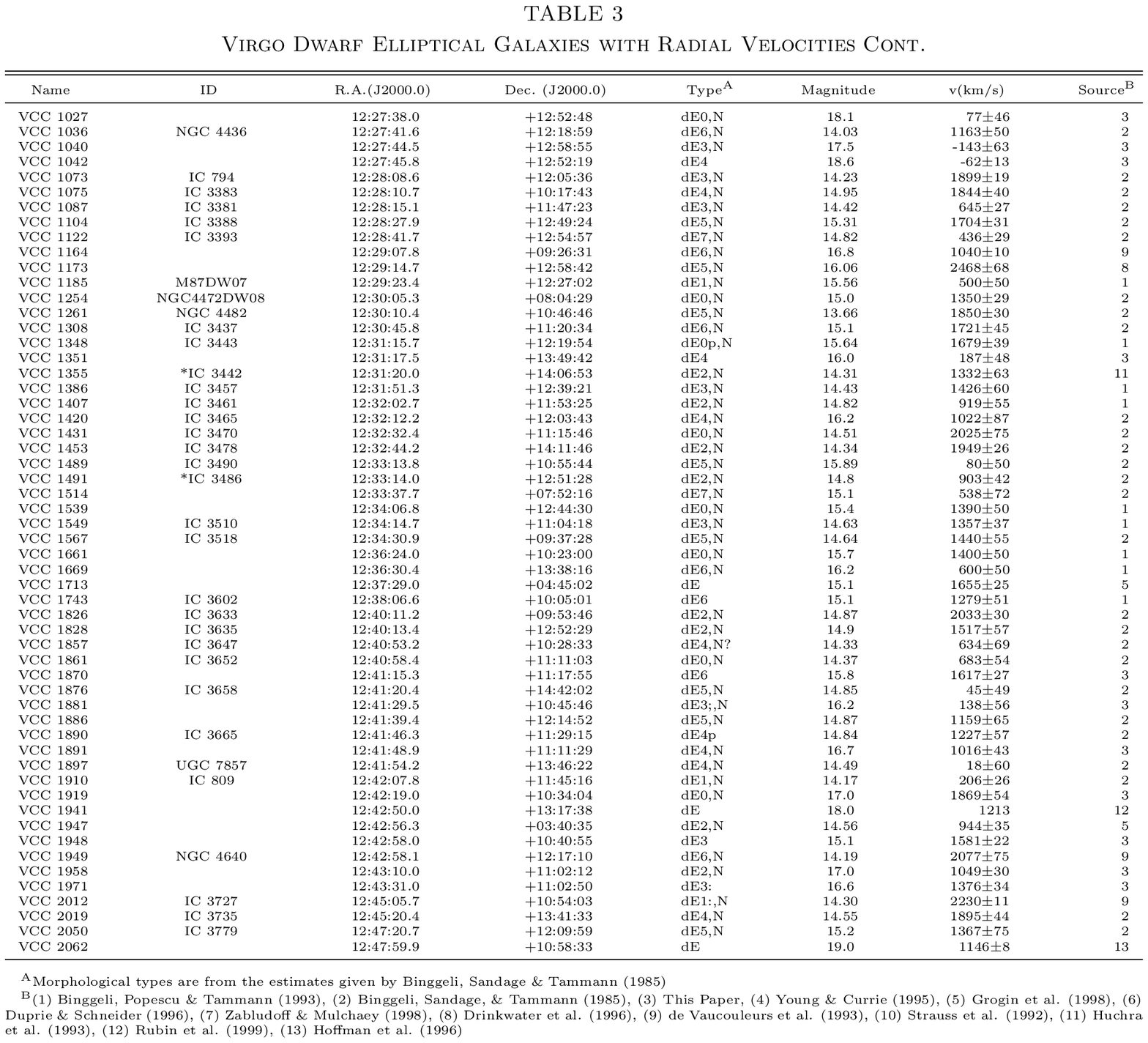}{6.0in}{0}{100}{100}{-310}{-170}
\end{figure}

\begin{figure}
\plotfiddle{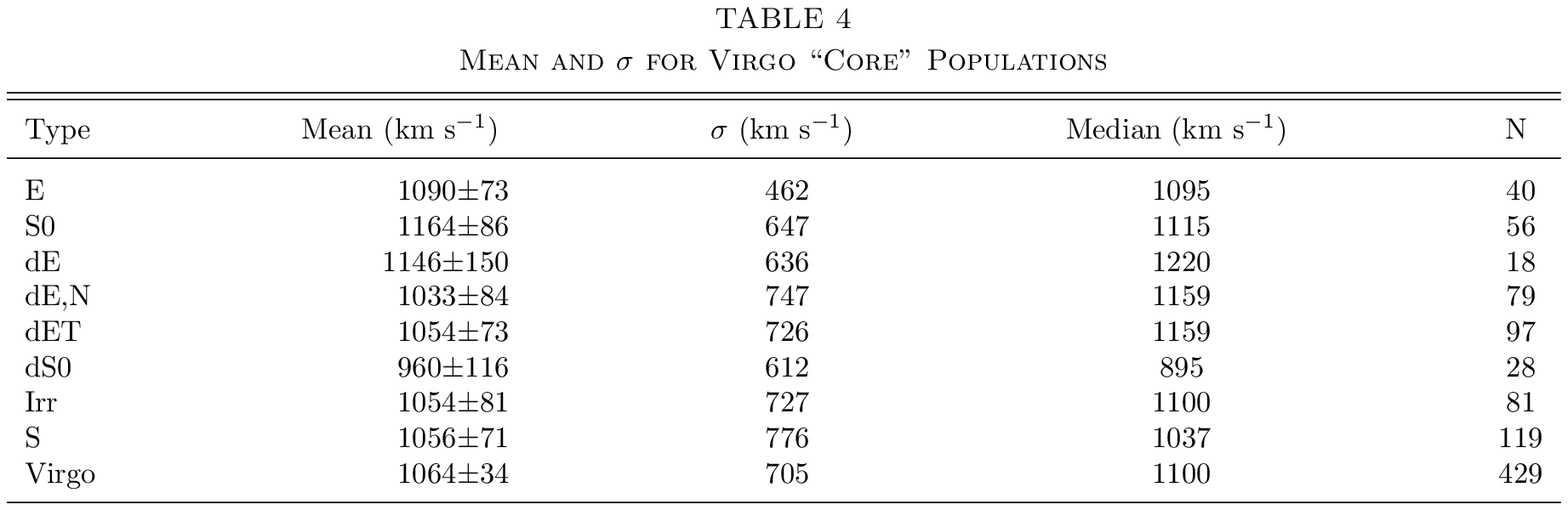}{6.0in}{0}{100}{100}{-310}{-170}
\end{figure}

\begin{figure}
\plotfiddle{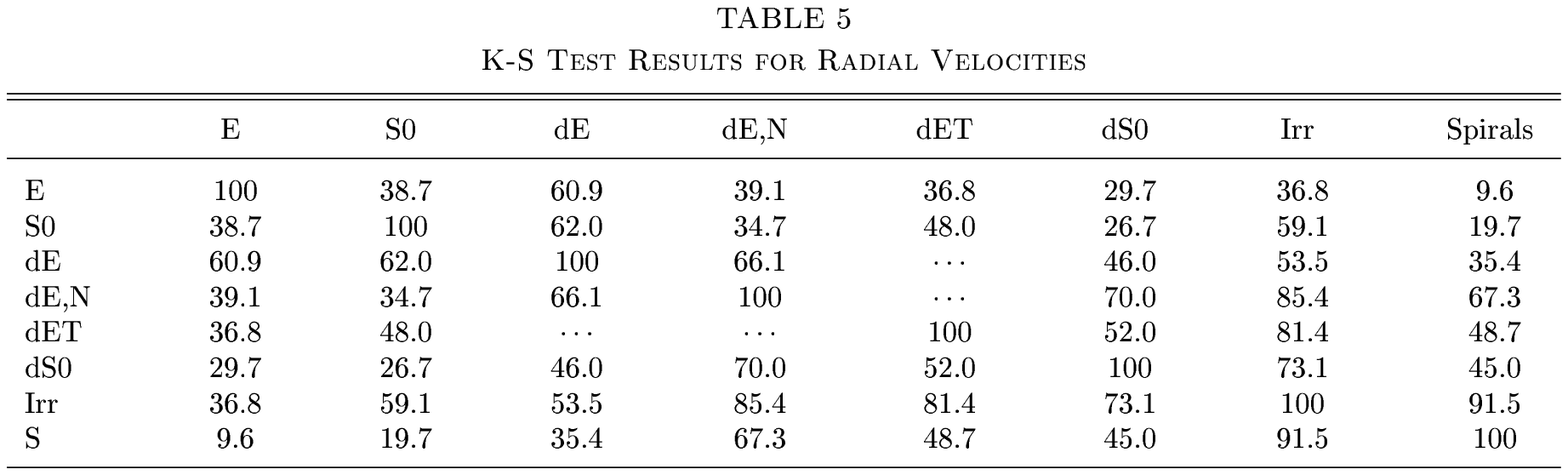}{6.0in}{0}{100}{100}{-310}{-170}
\end{figure}

\begin{figure}
\plotfiddle{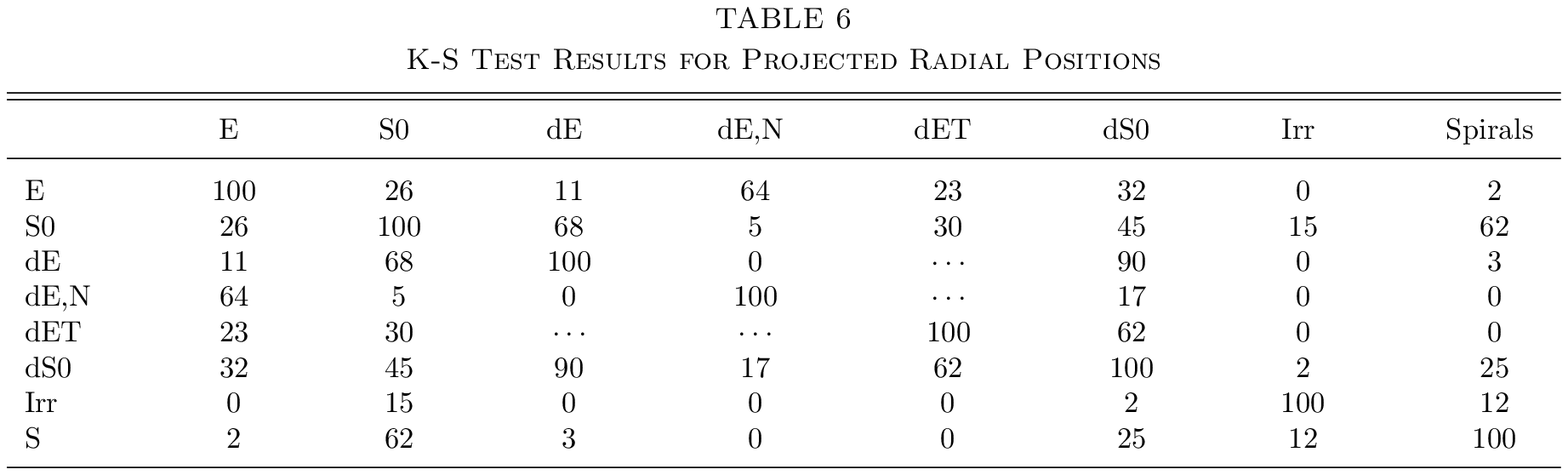}{6.0in}{0}{100}{100}{-310}{-170}
\end{figure}

\begin{figure}
\plotfiddle{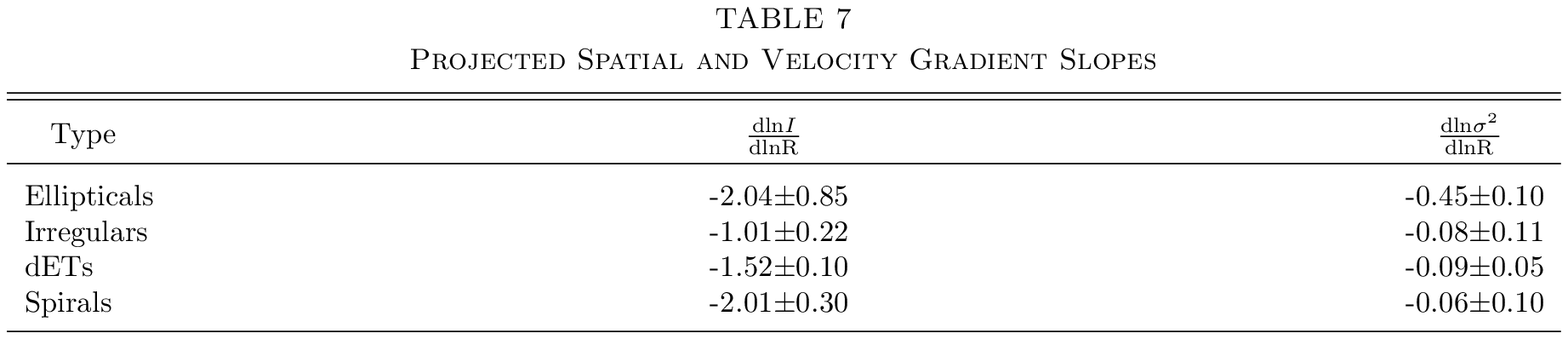}{6.0in}{0}{100}{100}{-310}{-170}
\end{figure}

%% file: figures_input.tex
\setcounter{figure}{0}

\clearpage
\begin{figure}
\plotfiddle{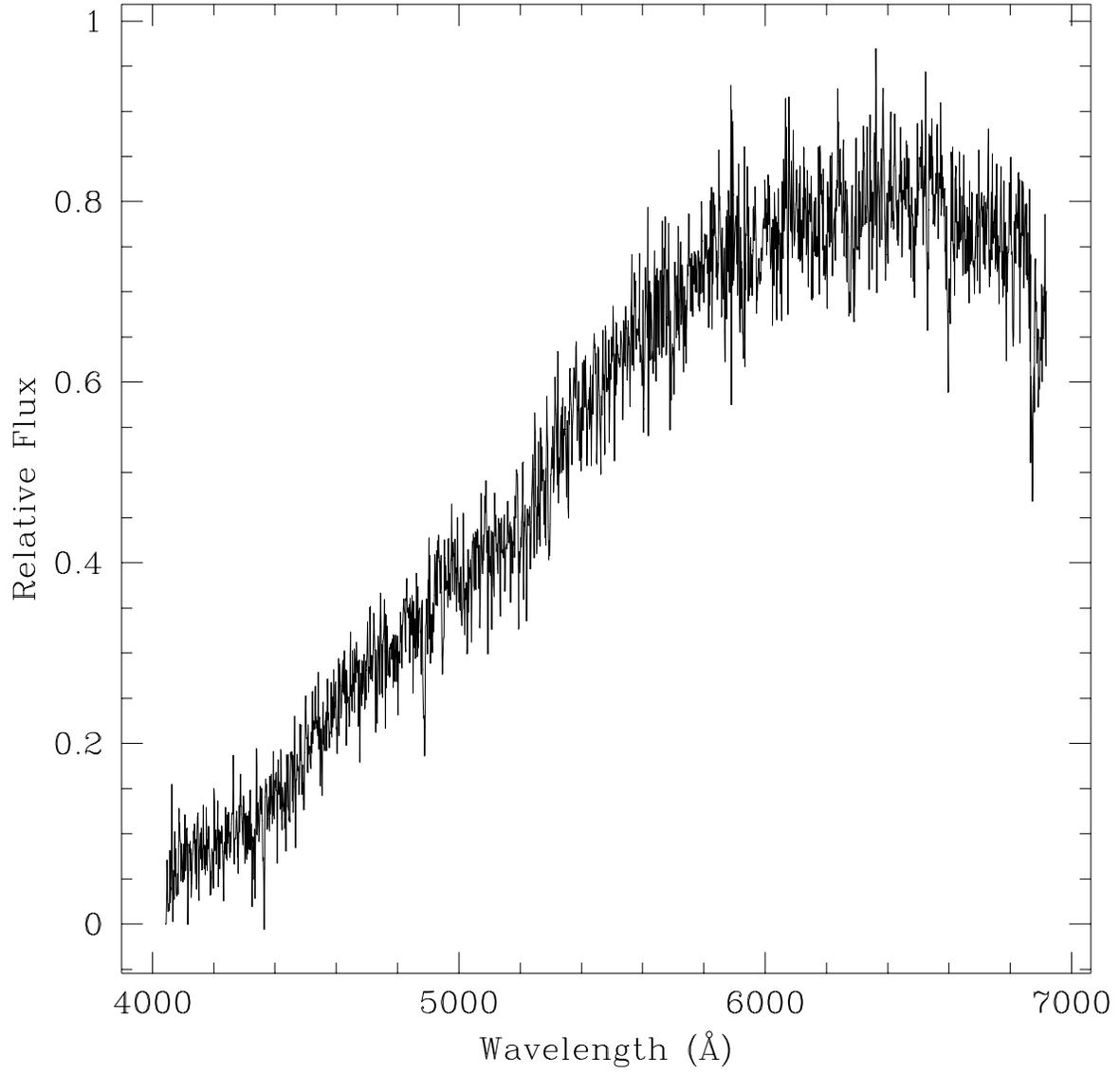}{6.0in}{0}{80}{80}{-250}{-100}
\caption{A typical WIYN Hydra Multi-Object Spectrograph spectrum of
a dwarf elliptical.}
\end{figure}

\clearpage
\begin{figure} 
\plotfiddle{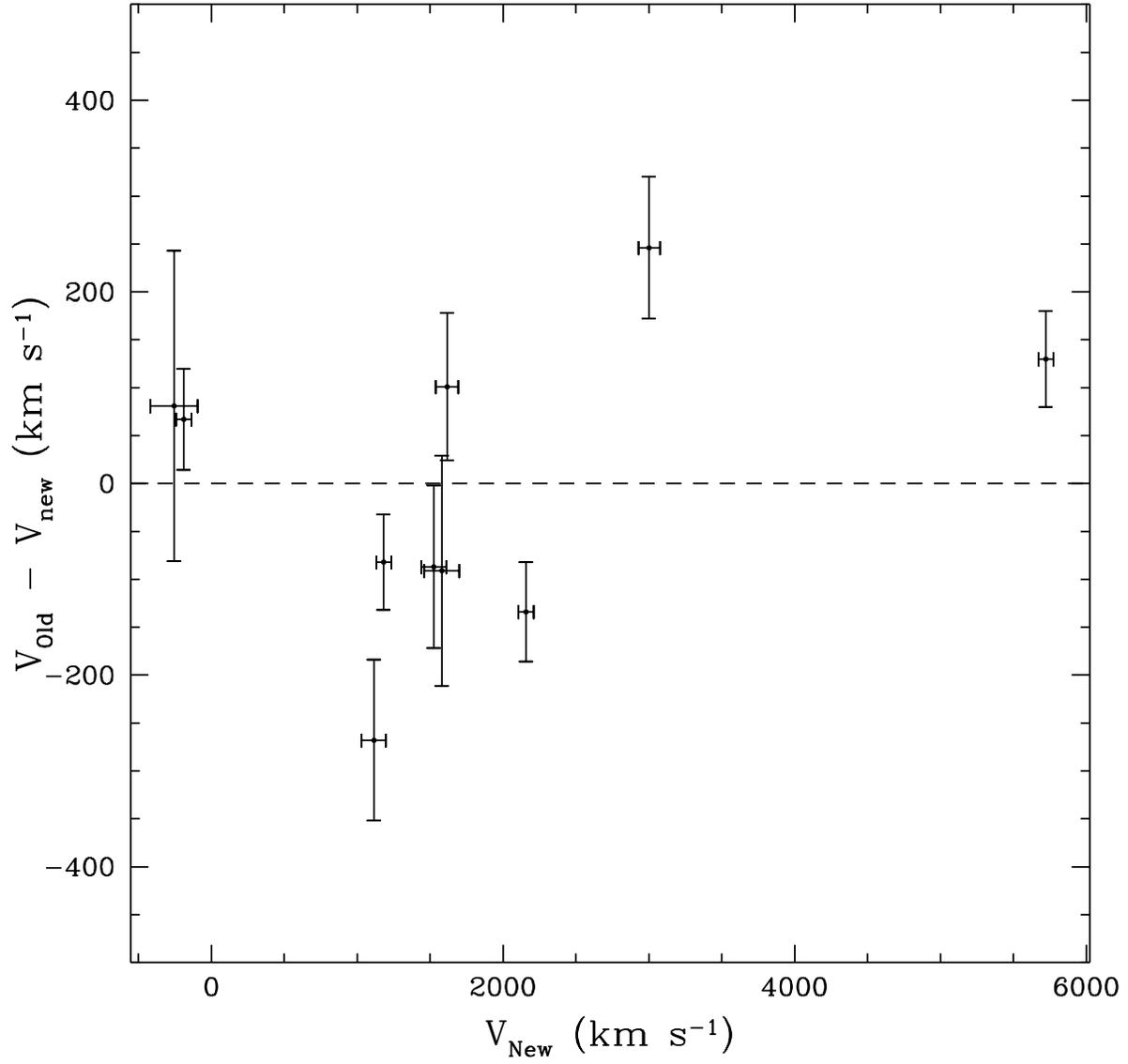}{6.0in}{0}{80}{80}{-250}{-100}
\caption{The difference between previously derived radial
velocities and values measured in this study.}
\end{figure}

\clearpage

\begin{figure}
\plotfiddle{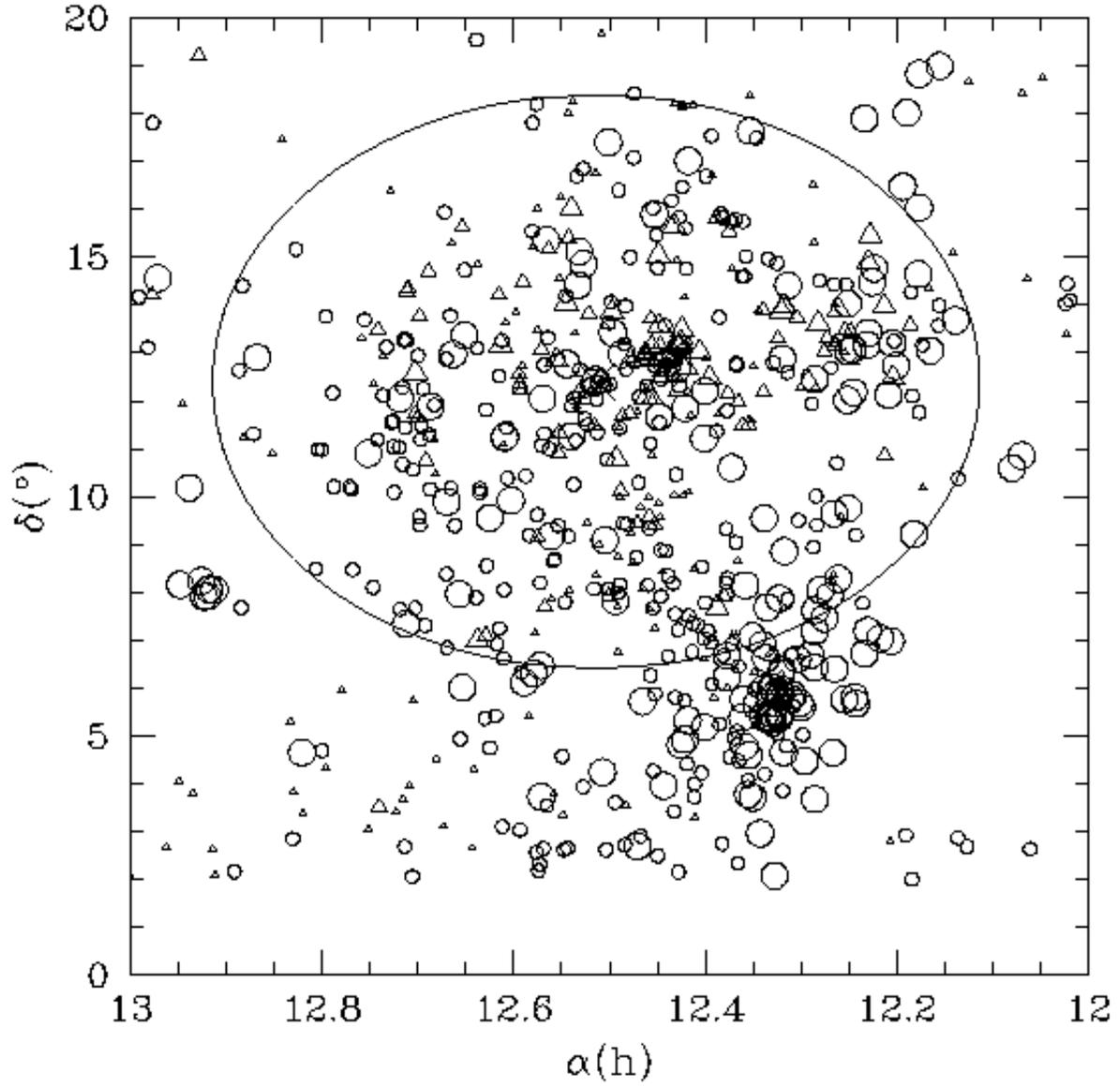}{6.0in}{0}{80}{80}{-250}{-100}
\vskip 0.2in
\caption{The projected distribution (J2000) of Virgo galaxies with 
known radial velocities.  The large X symbol represents the location of
the large elliptical M87.  The 6\deg circle represents the spatial
area used to define the Virgo core sample.  The different points for
galaxies represent different velocities.  The larger the symbol, the
larger the difference from 1000 \kms, with triangles for V$<1000$ \kms, and
circles for V$>1000$ \kms.}
\end{figure}

\clearpage

\begin{figure}
\plotfiddle{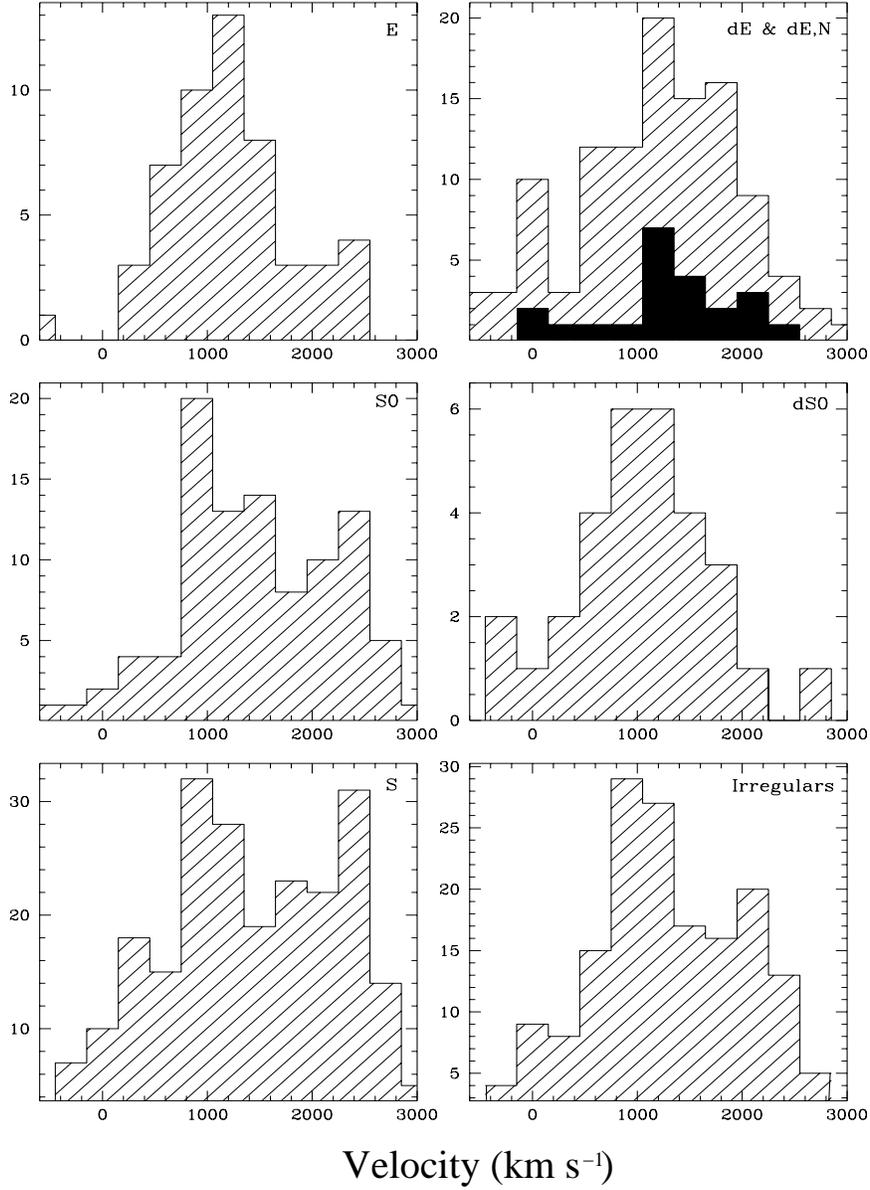}{6.0in}{0}{80}{80}{-250}{-100}
\caption{Velocity histograms of six galaxy populations contained in
the entire Virgo sample.  The solid part of the dE and dE,N histogram
represents the non-nucleated dEs, while the shaded area is the combined dE 
and dE,Ns.}
\end{figure}

\clearpage

\begin{figure}
\plotfiddle{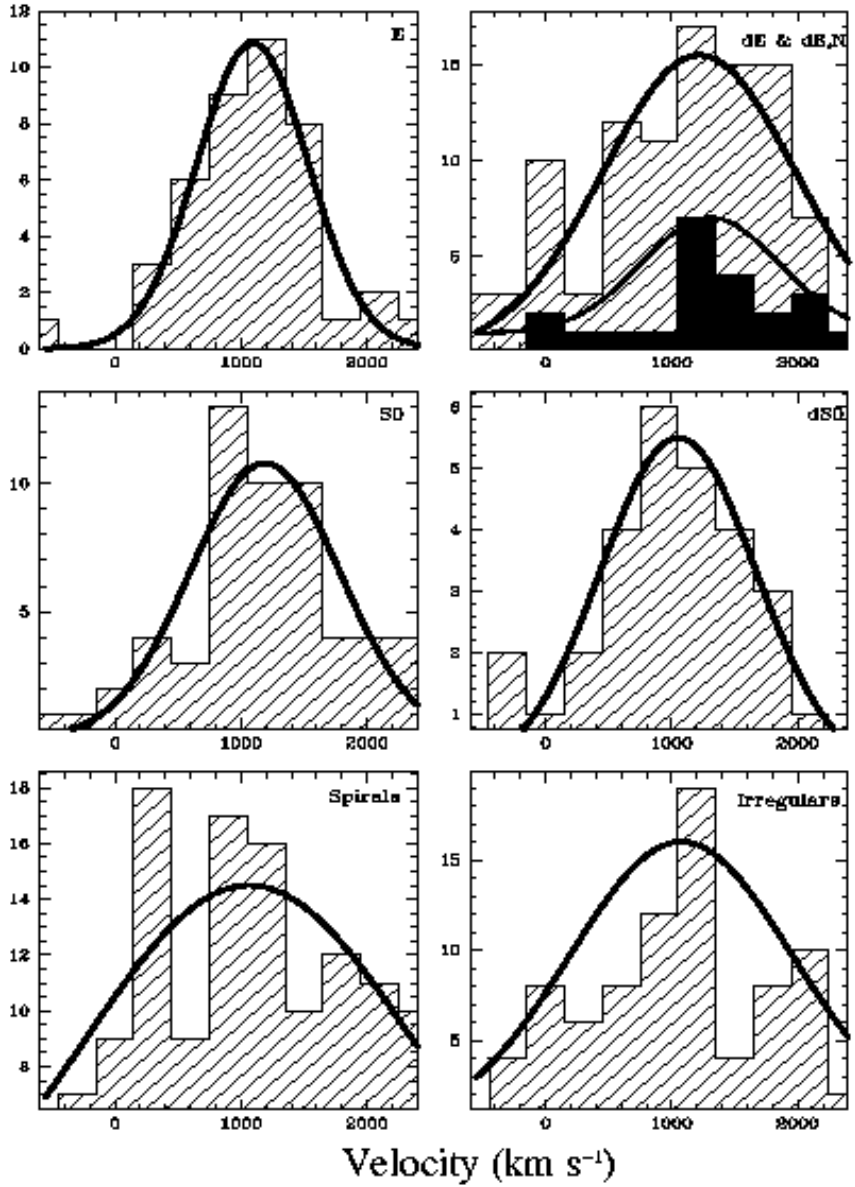}{7.0in}{0}{80}{80}{-250}{-100}
\vskip 0.2in
\caption{Velocity histograms for the Virgo `Core' population,
with fitted Gaussian profiles.  The galaxies here are the subsample
of galaxies in Figure 4 that are within 6\deg\, from the dynamical
center of Virgo (see text) and whose radial velocities are $< 2400$ \kms. 
The solid part of the dET histogram represents the pure dEs.}
\end{figure}

\clearpage

\begin{figure}
\plotfiddle{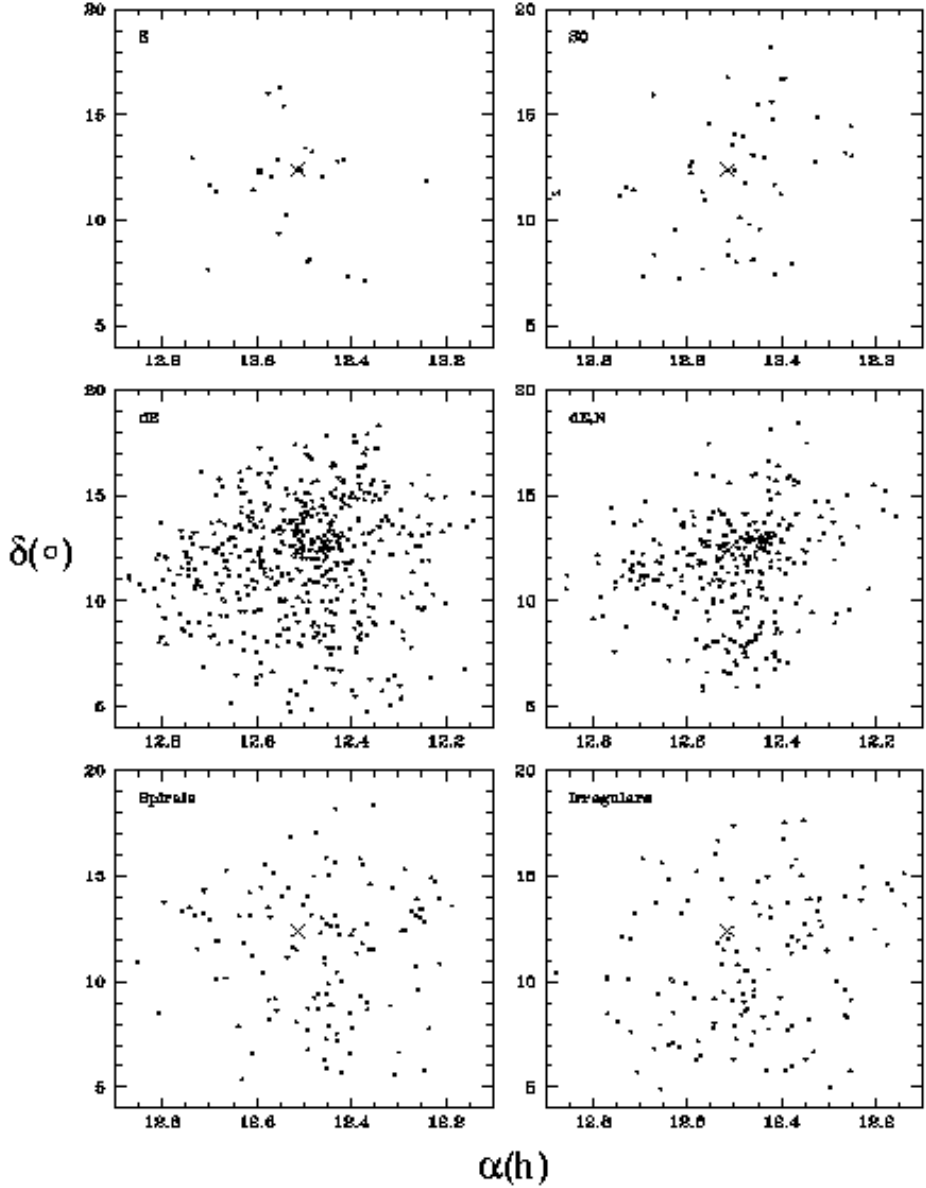}{7.0in}{0}{80}{80}{-250}{-100}
\vskip 0.2in
\caption{Positional plots of the various Virgo populations.  These
contain all galaxies considered by Binggeli et al. (1985) to be within
Virgo based on velocity and/or morphological information.  The X symbol
represents the location of M87.}
\end{figure}
\clearpage

\begin{figure}
\plotfiddle{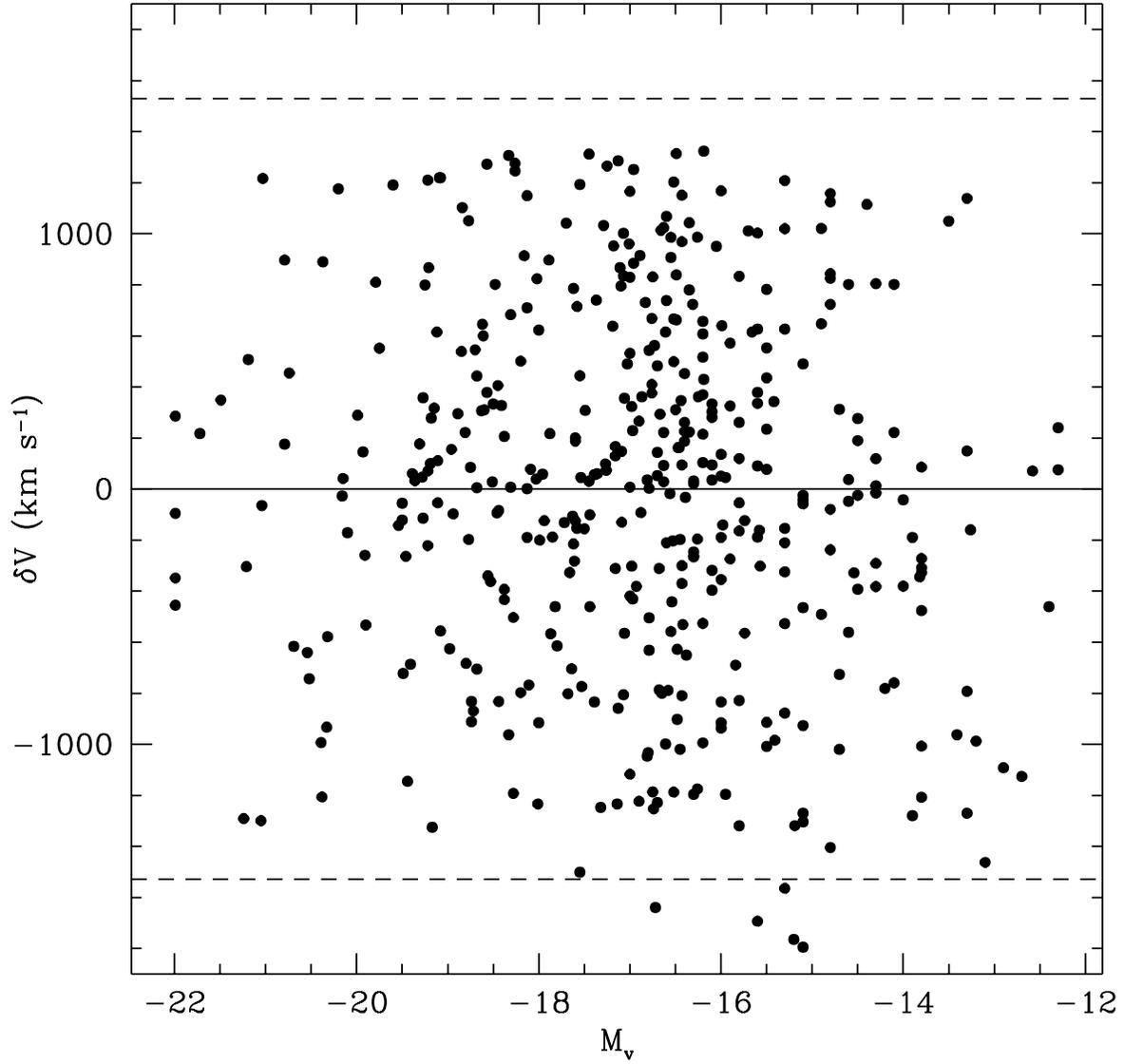}{7.0in}{0}{80}{80}{-250}{-100}
\vskip -0.5in
\caption{Relative velocities of Virgo galaxies plotted   
as a function of their absolute magnitudes assuming a Virgo
distance modulus of (m-M) = 31.3.  From this plot there is
no sign of mass-velocity segregation or energy equipartition in Virgo. The 
dashed lines represent the escape velocity of Virgo.}
\end{figure}
\clearpage

\begin{figure}
\plotfiddle{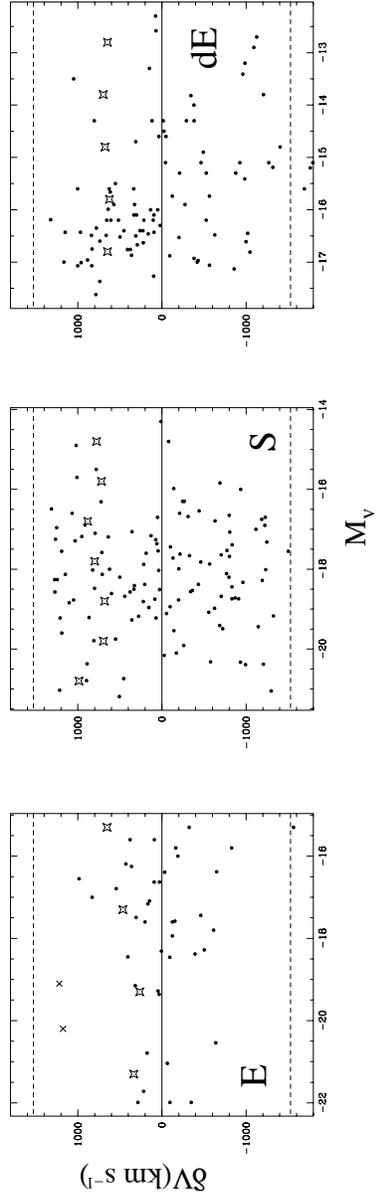}{7.0in}{0}{60}{60}{-190}{0}
\caption{Elliptical, spiral and dET velocities as a function of 
magnitude.   The open stars 
represent the velocity 
dispersion as a function of magnitude.  The two upper X symbols in the
elliptical galaxy plot are two galaxies in our sample that are likely not 
part of the Virgo cluster proper or true ellipticals (see text).}
\end{figure}

\clearpage

\begin{figure}
\plotfiddle{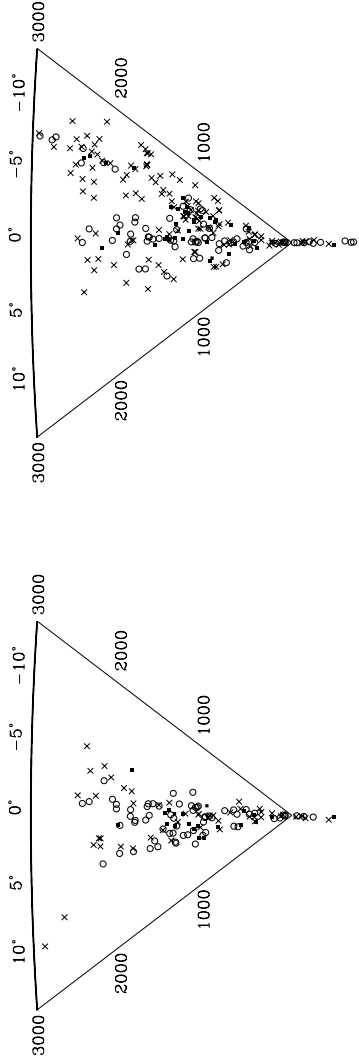}{6.0in}{0}{60}{60}{-190}{0}
\vskip -0.5in
\caption{Declination and right ascension velocity slices through the 
core of Virgo showing
the 'finger of god' signature and infall ring patterns.  The spirals
are X symbols, ellipticals black squares and dwarf ellipticals
open circles.  Several other clumpy regions can be seen in this
diagram, most particularly at (-3\deg, 1000 \kms) and (-5\deg, 2000 \kms).}
\end{figure}

\clearpage

\begin{figure}
\plotfiddle{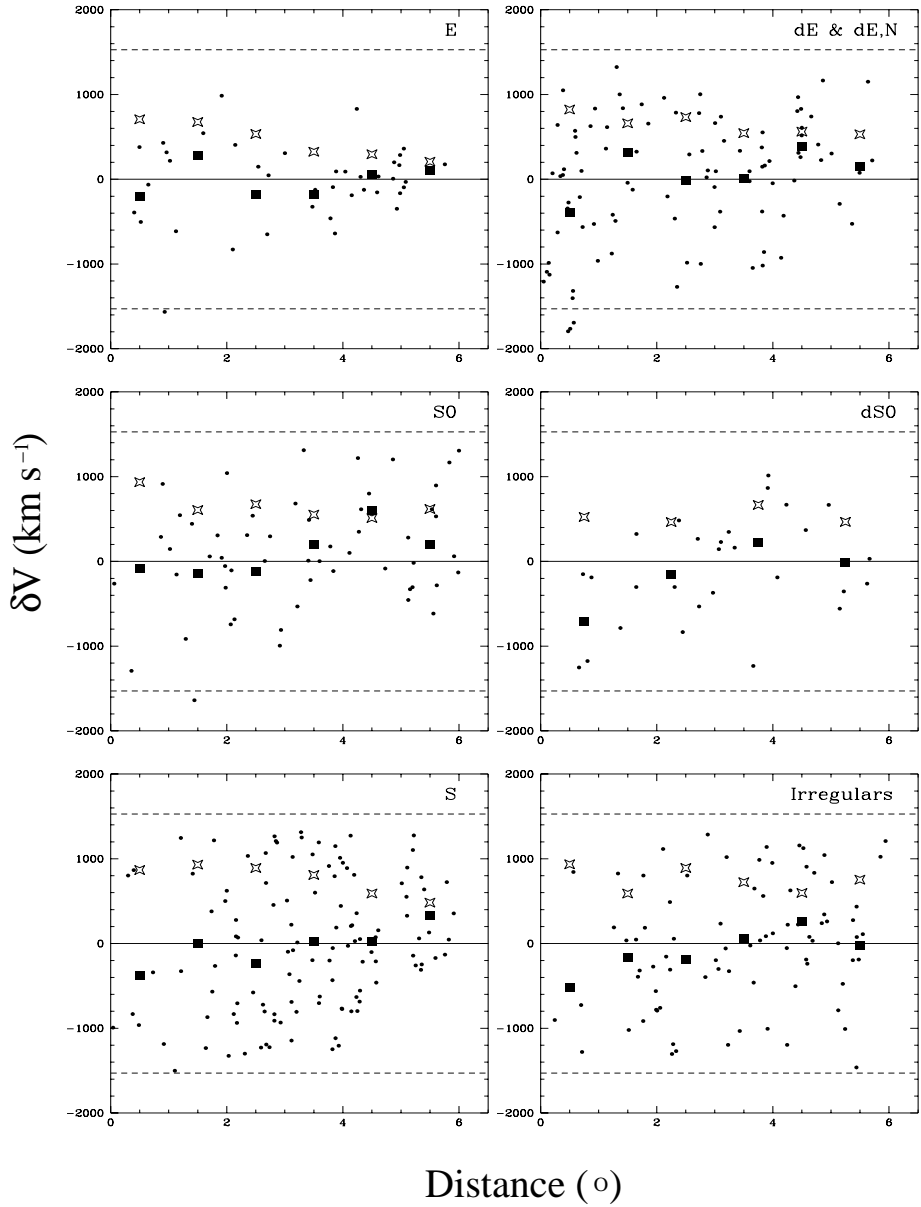}{7.0in}{0}{80}{80}{-250}{-100}
\caption{The radial velocities of the Virgo galaxy populations
plotted as function of projected distance from the center of Virgo.  The high
velocity dispersions (open stars) at the center of Virgo indicate
that galaxies in Virgo are on radial or highly elliptical orbits. The
black squares represent the average velocity at each distance.}
\end{figure}

\begin{figure}
\plotfiddle{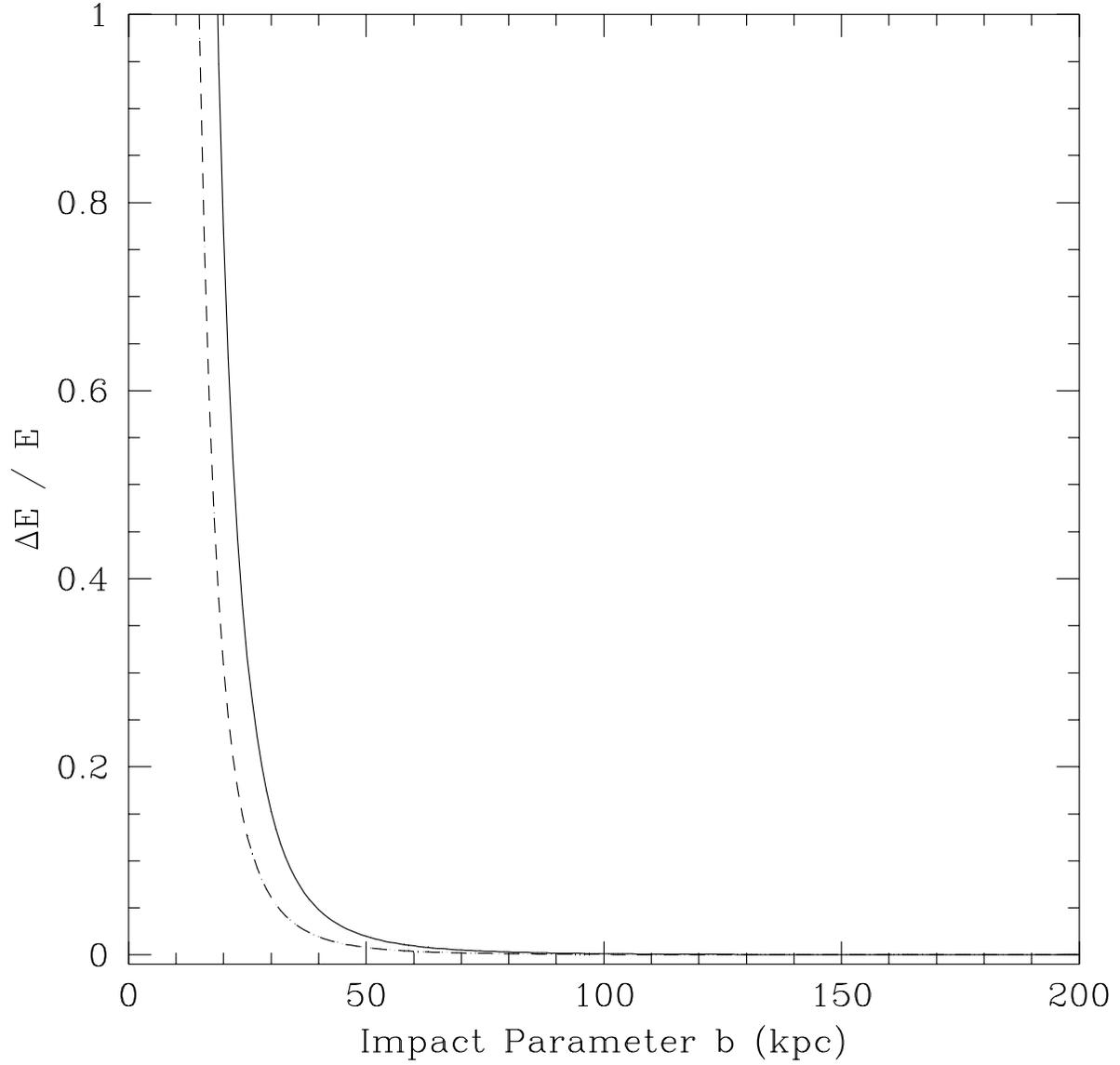}{7.0in}{0}{80}{80}{-250}{-100}
\vskip -0.1in
\caption{The ratio of the increase in internal energy produced in
an impulsive interaction to
the total internal energy, for two model galaxies as a function of
the impact parameter b.  For both a galaxy with a large size (5 kpc: solid 
line) and a smaller size (2 kpc: dashed line)  the increase in internal
energy is not enough to destroy the galaxy with impact parameters greater
than $\sim$ 30 kpc.}
\end{figure}

\clearpage

\begin{figure}
\plotfiddle{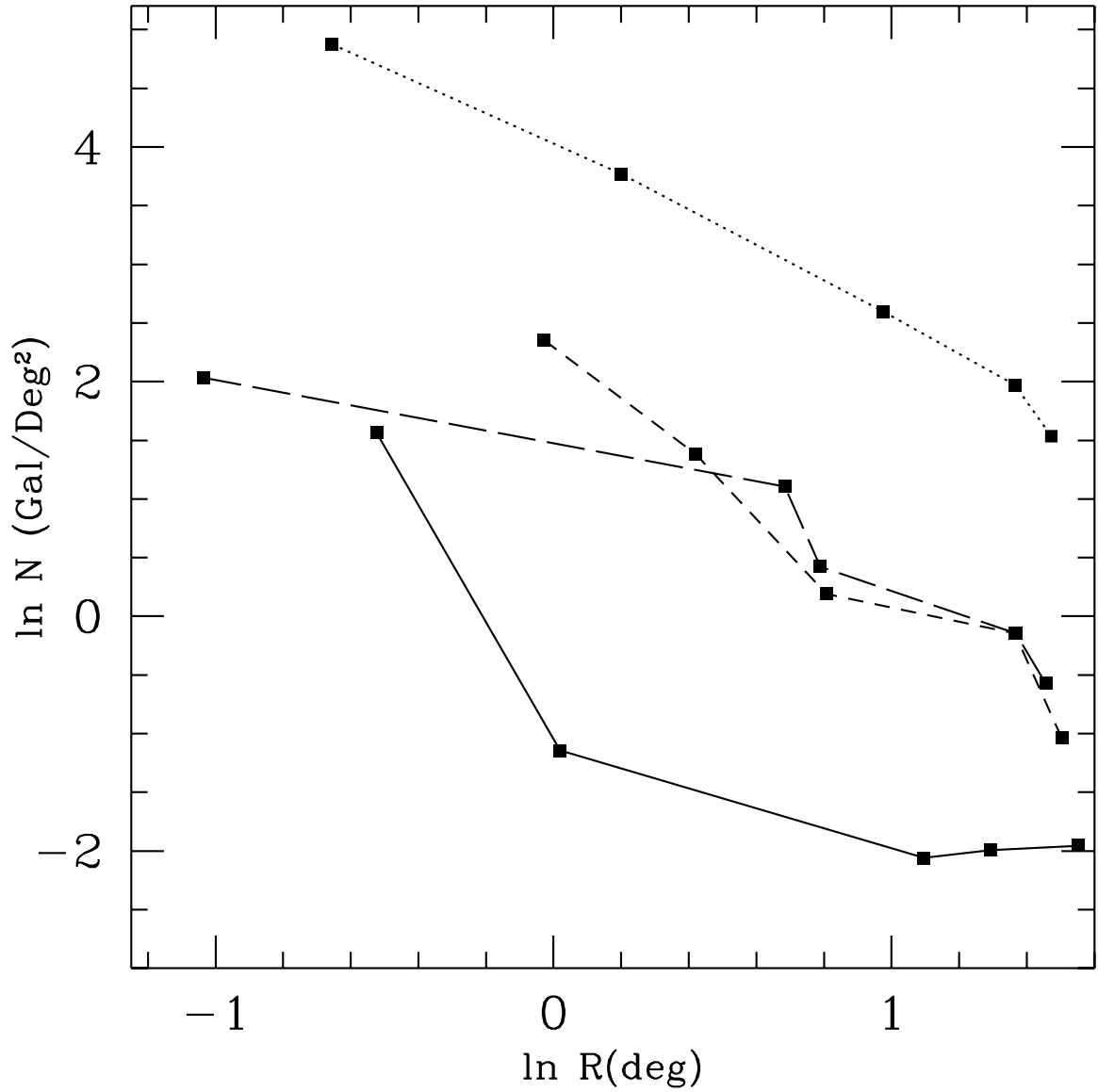}{7.0in}{0}{80}{80}{-250}{-100}
\caption{The projected spatial distribution, $N$, in galaxies per square 
degree as a function of radius.  Ellipticals have the lowest densities and are
represented by a solid line.  The spirals are short-dashed lines and irregulars
long-dashed ones.  The dETs, the densest population, are designated by the
small dotted line.}
\end{figure}